\documentclass[twocolumn, 10pt, pra, aps, showpacs, reprint, amsmath, amssymb, superscriptaddress, nofootinbib, floatfix]{revtex4-1}

\usepackage{times}
\usepackage[pdftex]{graphicx} % Include figure files

\begin{document}

\title{Massively parallel ultrafast random bit generation with a chip-scale laser}

\author{Kyungduk Kim}
\affiliation{Department of Applied Physics, Yale University, New Haven, Connecticut 06520, USA}
\author{Stefan Bittner}
\affiliation{Department of Applied Physics, Yale University, New Haven, Connecticut 06520, USA}
\affiliation{Chair in Photonics, LMOPS EA-4423 Laboratory, CentraleSup\'elec and Universit\'e de Lorraine, 2 rue Edouard Belin, Metz 57070, France}
\author{Yongquan Zeng}
\affiliation{Center for OptoElectronics and Biophotonics, School of Electrical and Electronic Engineering \& School of Physical and Mathematical Science and The Photonics Institute, Nanyang Technological University, 50 Nanyang Avenue, 639798 Singapore}
\author{Stefano Guazzotti}
\author{Ortwin Hess}
\affiliation{The Blackett Laboratory, Imperial College London, London SW7 2AZ, United Kingdom}
\affiliation{School of Physics and CRANN Institute, Trinity College Dublin, Dublin 2, Ireland}
\author{Qi Jie Wang}
\affiliation{Center for OptoElectronics and Biophotonics, School of Electrical and Electronic Engineering \& School of Physical and Mathematical Science and The Photonics Institute, Nanyang Technological University, 50 Nanyang Avenue, 639798 Singapore}
\author{Hui Cao}
\email{hui.cao@yale.edu}
\affiliation{Department of Applied Physics, Yale University, New Haven, Connecticut 06520, USA}

\begin{abstract}
Random numbers are widely used for information security, cryptography, stochastic modeling, and quantum simulations. Key technical challenges for physical random number generation are speed and scalability. We demonstrate a method for ultrafast generation of hundreds of random bit streams in parallel with a single laser diode. Spatio-temporal interference of many lasing modes in a specially designed cavity is introduced as a scheme for greatly accelerated random bit generation. Spontaneous emission, caused by quantum fluctuations, produces stochastic noise that makes the bit streams unpredictable. We achieve a total bit rate of 250 terabits per second with off-line post-processing, which is more than two orders of magnitude higher than the current post-processing record. Our approach is robust, compact, and energy efficient with potential applications in secure communication and high-performance computation.
\end{abstract}

\maketitle

% Figure 1
\begin{figure*}[t]
\begin{center}
\includegraphics[width = 0.8\linewidth]{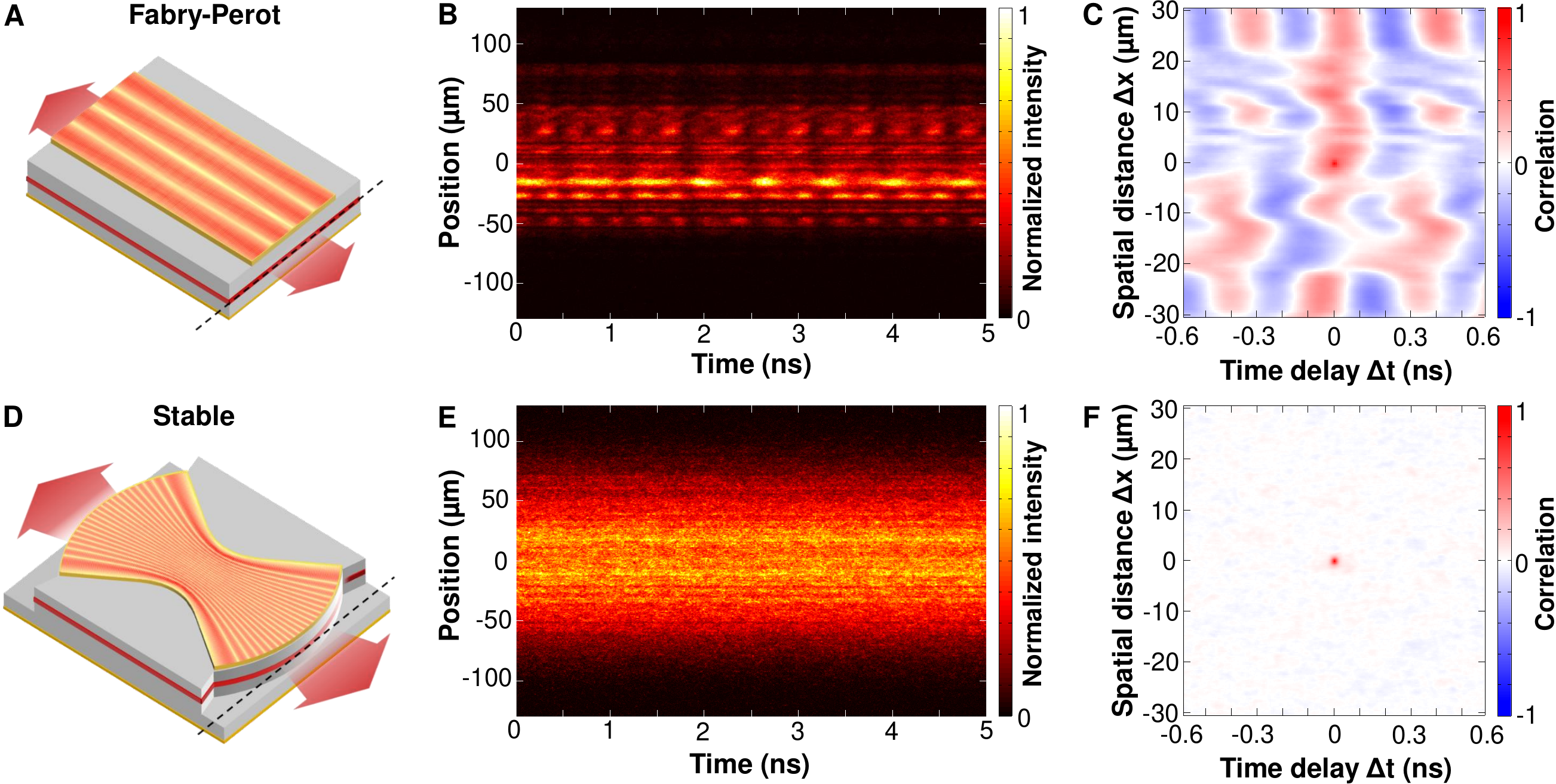}
\end{center}
\caption{Reducing spatio-temporal correlations of the lasing emission.
    (\textbf{A})~A wide-stripe edge-emitting semiconductor laser with planar facets supports only low-order transverse modes with the typical spatial profile shown (not to scale). 
	(\textbf{B})~Emission intensity $I(x,t)$ at one facet of a 100-$\mu$m-wide, 1000-$\mu$m-long GaAs quantum-well laser, measured by a streak camera, features filamentation and irregular pulsations.
	(\textbf{C})~The spatio-temporal correlation function $C(\Delta x,\Delta t)$ of the emission intensity in (B) reveals long-range spatio-temporal correlations. 
	(\textbf{D})~Our specially-designed laser cavity with curved facets confines high-order transverse modes. The spatial intensity distribution of an exemplary high-order transverse mode is plotted.
	(\textbf{E})~The measured spatio-temporal trace of the lasing emission from our cavity of length 400~$\mu$m, width 282~$\mu$m and facet radius 230~$\mu$m is free of micrometer-sized filaments and GHz oscillations as seen in (B). 
	(\textbf{F})~The spatio-temporal correlation function $C(\Delta x,\Delta t)$ of the emission intensity in (E) shows no long-range spatio-temporal correlations.
}
\label{fig1}
\end{figure*}

% Figure 2
\begin{figure}[t]
\begin{center}
\includegraphics[width = \linewidth]{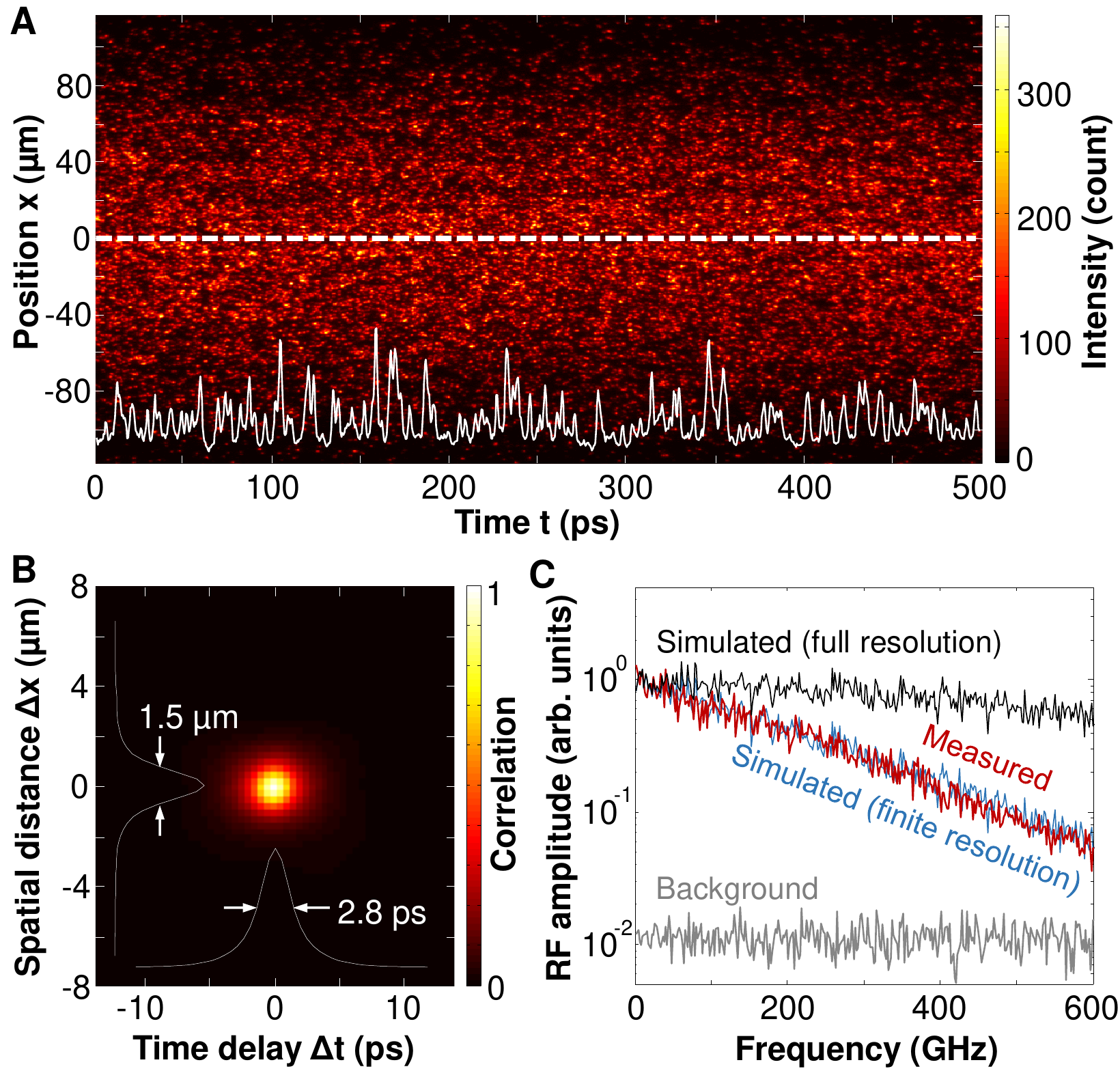}
\end{center}
\caption{Ultrafast beating of lasing modes.
    (\textbf{A})~The lasing emission at one facet of our 600-$\mu$m-long cavity exhibits a spatio-temporal interference pattern.  
	The white solid curve is the temporal intensity fluctuation at position $x$ = 0 (white dashed line). 
	(\textbf{B})~The correlation function $C(\Delta x,\Delta t)$ of the measured lasing emission intensity in (A) gives the spatial and temporal correlation widths of 1.5~$\mu$m and 2.8~ps. 
	(\textbf{C})~The RF spectrum (the modulus of Fourier transform) of the emission intensity in (A) at $x=0$ (red) is much higher than the background (gray) with the laser turned off. The simulated spectrum (black) is broader, but becomes narrower when the temporal resolution of our detector is taken into account (blue), in agreement with the measured one (red).
}
\label{fig2}
\end{figure}

% Figure 3
\begin{figure*}[t]
\begin{center}
\includegraphics[width = 0.9\linewidth]{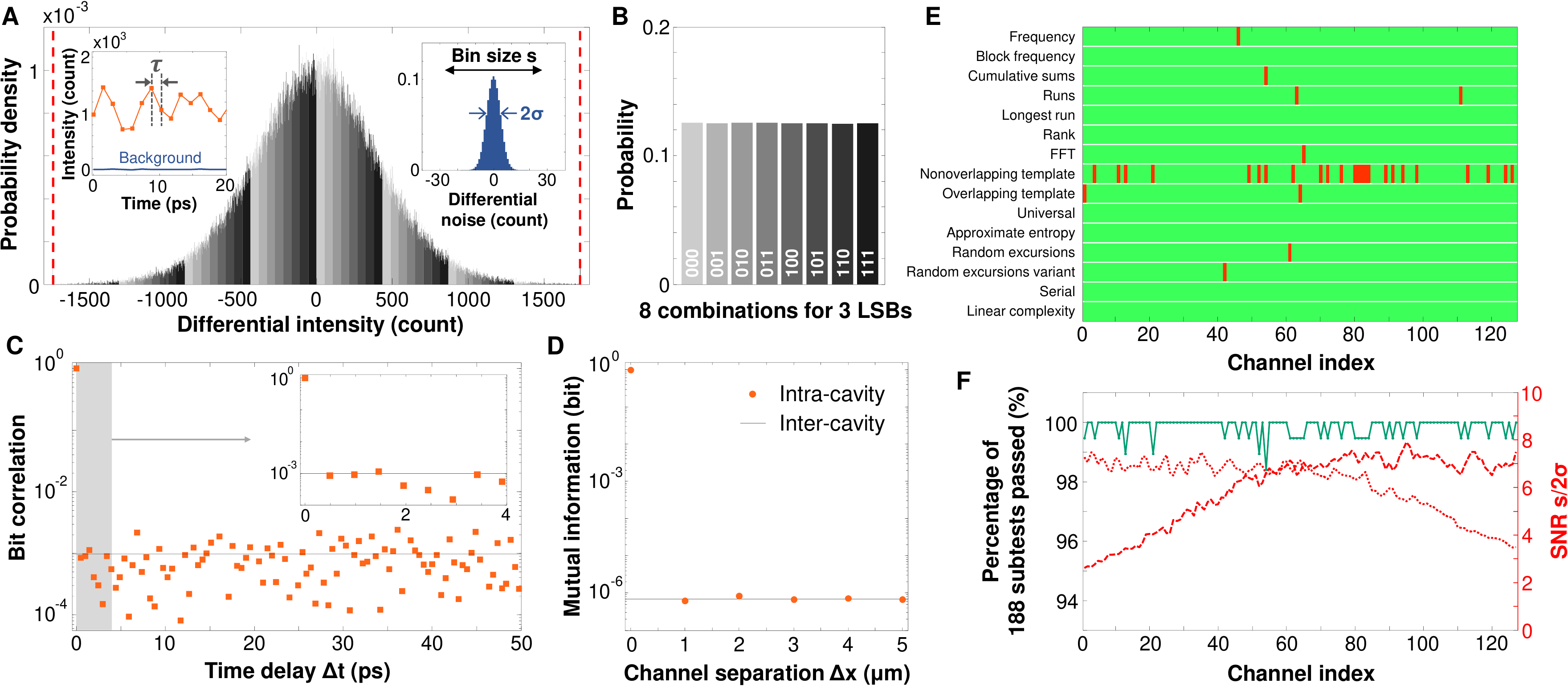}
\end{center}
\caption{Parallel random bit generation and evaluation.
	(\textbf{A})~The PDF of the differential intensity, $\Delta I_n = I_{n+4} - I_n$, which is digitized to 6 bits by binning the range [-1740, 1740] counts (vertical red dashed lines) into 64 equally spaced intervals. Three LSBs are taken from each sample. The gray scale of the bars represents their eight combinations. Left inset: A segment of intensity time trace of a single spatial channel (red line), sampled at intervals $\tau$ = 1.46~ps (red dots). The blue curve is the background count. Right inset: The PDF of the differential background count with a standard deviation $\sigma = 3.9$ counts, much smaller than the bin size $s = 54$ for $\Delta I_n$.
	(\textbf{B})~The probability for all eight combinations of three LSBs is almost equal. 
	(\textbf{C})~A bit stream with length $N$ = $2^{20}$ has a bit correlation (red squares) around the lower limit $1/\sqrt{N}$ (black line). Inset: Close-up of short delay times. 
	(\textbf{D})~The mutual information between bit streams in two channels with varying separation (Intracavity) is equal to that between the streams from two independent lasers (Intercavity).
    (\textbf{E})~The NIST SP800-22 test results include 15 kinds of statistical tests, yielding a total of 188 subtests for 127 parallel bit streams. The green color denotes one stream passing one test; conversely, red denotes test failure. Ninety-five bit streams pass all subtests, yielding a pass rate of 75\%, which is considered acceptable for a reliable RBG.
    (\textbf{F})~The percentage of all 188 subtests that every bit stream passes (green) is uncorrelated with the SNR $s/2\sigma$ of the corresponding pair of spatial channels (red dashed and dotted lines).}
\label{fig3}
\end{figure*}

% Figure 4
\begin{figure}[b]
\begin{center}
\includegraphics[width = \linewidth]{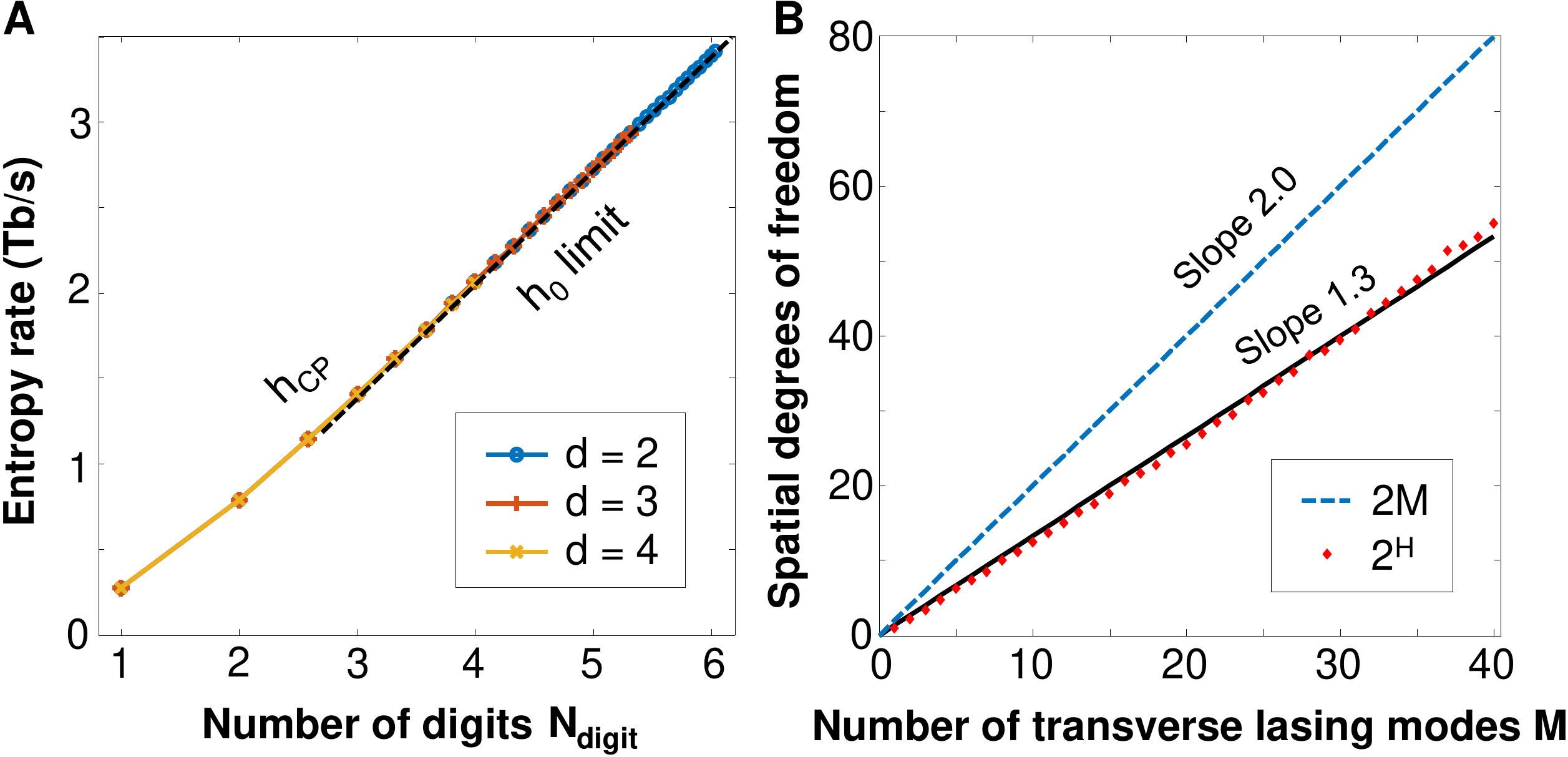}
\end{center}
\caption{Information capacity of spatio-temporal intensity pattern of simulated lasing emission.
    (\textbf{A})~The entropy rate $h_\mathrm{CP}$ in a single spatial channel converges for different embedding dimensions $d$, and reaches the information theoretical limit $h_0$ (black dashed line), indicating that the maximum possible RBG rate is reached.
	(\textbf{B})~The effective number of spatial degrees of freedom $2^H$ in the emission pattern grows linearly with the number of transverse lasing modes $M$. The black solid line is a linear fit. The blue dashed line is the naively expected number of spatial degrees of freedom $2M$, which exceeds $2^H$.
}
\label{fig4}
\end{figure}

The performance and reliability of our digital networked society are based on the ability to generate large quantities of randomness. An ever-increasing demand to improve the security of digital information has shifted the generation of random numbers from sole reliance on pseudo-random algorithms to the use of physical entropy sources. Ultrafast physical random number generators are key devices for achieving ultimate performance and reliability in communication and computation systems \cite{hart2017recommendations, herrero2017quantum}. Semiconductor lasers that feature chaotic dynamics with tens-of-GHz bandwidth represent one prominent class of high-speed random number generators \cite{uchida2008fast, reidler2009ultrahigh, kanter2010optical, argyris2010implementation, hirano2010fast, zhang2012robust, li2013heterodyne, oliver2013fast, fang2014noise, virte2014physical, sakuraba2015tb, tang2015tbits, butler2016optical, shinohara2017chaotic, ugajin2017real, xiang20192}. Initially, 1.7 Gb/s random bit generation (RBG) was achieved with combined binary digitization of two independent chaotic laser diodes \cite{uchida2008fast}. Then from a single chaotic semiconductor laser, a 12.5 Gb/s RBG was demonstrated \cite{reidler2009ultrahigh} with a subsequent boost to 300 Gb/s \cite{kanter2010optical}. By coupling several lasers to further increase the bandwidth and using post-processing schemes to extract more bits in analog-to-digital conversion (ADC), the total RBG rate was pushed up to 2 Tb/s \cite{sakuraba2015tb, tang2015tbits, butler2016optical, xiang20192}. However, the intrinsic time scales of lasing instabilities impose an ultimate limit on the entropy generation rate. A further increase in the RBG rate requires a different physical process with inherently faster dynamics.

Parallel RBG schemes can greatly enhance the generation rate and the scalability by producing many bit streams simultaneously. In the spatial domain, parallel generation of physical random numbers was realized by sampling two-dimensional laser speckle patterns created by a moving diffuser or a vibrating multimode fiber \cite{marron1986generation, lalanne19902}. As a result of inherently long mechanical timescales, the generation rates remain low (Mb/s). Chaotic broad-area semiconductor lasers have been investigated for high-speed parallel-RBG \cite{arahata2015inphase}, but correlations of intensity fluctuations at different spatial locations impede independent parallel bit stream generation. Spectral demultiplexing of amplified spontaneous emission \cite{li2011scalable,li2019parallel} or heterodyning chaotic laser emission \cite{li2013heterodyne} are used for parallel RBG with rates up to hundreds of Gb/s per channel. So far, such spectral-domain parallel RBG has been demonstrated with fewer than 10 channels.

We demonstrate a method that enhances the random bit rate in a single channel and also provides hundreds of channels for simultaneous generation of independent bit streams. The spatio-temporal interference of many lasing modes is used to generate picosecond-scale emission intensity fluctuations in space, so as to massively produce ultrafast random bit streams in parallel. This is achieved by tailoring the geometry of a broad-area semiconductor laser to vastly increase the number of transverse lasing modes, thereby suppressing characteristic dynamical instabilities such as filamentation. Specifically, we have designed a chip-scale laser diode to enable a large number of spatial modes lasing simultaneously with incommensurate frequency spacings, so that their interference patterns are complex and aperiodic. Spontaneous emission adds stochastic noise to make the intensity fluctuations unpredictable and non-reproducible. 

A conventional broad-area edge-emitting semiconductor laser has a stripe geometry with two flat facets (Fig.~\ref{fig1}A). Characteristically, lasing occurs only in the low-order transverse modes. Nonlinear interactions between the light field and the gain material entail irregular pulsation and filamentation \cite{Hess1996semiconductor} (Fig.~\ref{fig1}B). The spatio-temporal correlation function of the intensity fluctuations \cite{fischer1996complex, sm} $C(\Delta x,\Delta t)$ reveals non-local correlations in space and time (Fig.~\ref{fig1}C). On one hand, long-range temporal correlation reflects memory which degrades the quality of random bits generated at one spatial location. On the other hand, long-range spatial correlation means that the random bit streams generated at different locations are not completely independent, thus impeding parallel RBG \cite{arahata2015inphase}.

To achieve massively parallel ultrafast RBG, we enhance the number of transverse lasing modes by increasing the cavity width and curving the end facets (Fig.~\ref{fig1}D), effectively suppressing modulational instabilities. High-order transverse modes are well confined inside such a cavity, and their optical gain is enhanced by tailoring the top metal contact shape \cite{sm}. The number of transverse lasing modes is maximized by fine-tuning the cavity geometry \cite{kim2019electrically}. Lasing on small length scales of transverse wavelengths of high-order modes prevents lensing and self-focusing effects \cite{bittner2018suppressing} that would normally cause filamentation and instabilities (Fig.~\ref{fig1}E). In turn, the absence of filaments and pulsations eliminates long-range spatio-temporal correlations in the lasing intensity (Fig.~\ref{fig1}F). It is this shortening of the correlation lengths in space and time that paves the ground for a substantial increase in the number of independent spatial channels for parallel RBG, as well as a great enhancement of the RBG rate of every individual spatial channel.   

With lasing instabilities suppressed, the dynamic variations of the emission intensity are orchestrated by the interference of lasing modes with different frequencies. The characteristic time scale of such intensity fluctuations is inversely proportional to the spectral width of the total emission, and is $\sim$1~ps for the GaAs quantum-well laser \cite{sm}. We show the spatio-temporal beat pattern of the intensity emitted at one laser facet in Fig.~\ref{fig2}A. The temporal correlation length is determined by the full width at half maximum (FWHM) of $C(\Delta x,\Delta t)$ in time (Fig.~\ref{fig2}B). Its value of 2.8 ps is limited by the temporal resolution of our detection \cite{sm}.

Thanks to the ultrafast dynamics of lasing intensity, the radio-frequency (RF) spectrum is extremely broad (Fig.~\ref{fig2}C). Its bandwidth, which contains 80$\%$ of the entire spectrum, is 315 GHz. For comparison, the numerically-simulated spectrum is even broader with a bandwidth of 632 GHz \cite{sm}. After accounting for the temporal resolution of photodetection, the simulated RF spectrum matches the measured one (Fig.~\ref{fig2}C). This agreement confirms that the ultra-broad spectrum results from the interference of many transverse and longitudinal modes.

For spatial-multiplexing of RBG, the number of independent parallel channels depends on the spatial correlation length of the lasing emission. Now with non-local correlations removed, the local correlation length estimated from the spatial FWHM of $C(\Delta x,\Delta t)$ is 1.5~$\mu$m (Fig.~\ref{fig2}B), which is limited by the spatial resolution of our detection. Without the finite experimental resolution, our simulation gives a correlation length of 0.5~$\mu$m \cite{sm}. Thanks to this extremely short spatial correlation length, hundreds of independent spatial channels are available for parallel RBG.

Because the transverse mode frequency spacing in our cavity design is incommensurate to the longitudinal mode spacing, the spatio-temporal interference pattern cannot repeat itself \cite{sm}. Moreover, the spontaneous emission, generated by quantum fluctuations, constantly feeds stochastic noise into the lasing modes, making their beat pattern unpredictable and irreproducible. 

To generate random bits, we divide the laser end facet into 1-$\mu$m-wide spatial channels. Because of the restricted field of view of our collection optics, only 254 spatial channels are recorded simultaneously, which is about half of the number possible with complete collection of emission. We sample the emission intensity at every spatial channel at intervals of $\tau$ = 1.46~ps (sampling rate 683~GHz; Fig.~\ref{fig3}A, left inset) \cite{sm}. The emission intensity integrated over one sampling period $I_n$ has an asymmetric probability density function (PDF), which would yield biased bits \cite{sm}. We adopt the procedure from \cite{reidler2009ultrahigh} to acquire the differential intensity $\Delta I_n = I_{n+4} - I_n$, which has a symmetric PDF (Fig.~\ref{fig3}A). 
The differential intensity $\Delta I_n$ is digitized to 6 bits (Fig.~\ref{fig3}A), and three least significant bits (LSBs) are used for RBG \cite{reidler2009ultrahigh}. All eight combinations for three LSBs have almost equal probability (Fig.~\ref{fig3}B). We remove the residual bias by performing exclusive-OR (XOR) on two bit streams from distant spatial channels \cite{sm}, which reduces the number of parallel bit streams to 127. Figure~\ref{fig3}C reveals that the correlation between bits in a single bit stream reaches the limit $1/\sqrt{N}$ given by the bit stream length $N$. This leads to a single-channel bit generation rate of 2~Tb/s, which is twice the current single-channel record with off-line post-processing~\cite{sakuraba2015tb, tang2015tbits, butler2016optical}.

We evaluated the quality of the generated random bits with two standard statistical test suites: NIST SP 800-22 and Diehard \cite{sm}. Fig.~\ref{fig3}E provides the test results for 127 parallel bit streams; 95 of them passed all NIST tests, yielding a pass rate of 75\%. Compared to the pass rates reported previously for pseudo-RBG and physical RBG, our pass rate is within the acceptable range for reliable random bit generators \cite{lihua2014study, shinohara2017chaotic}. In addition, we performed the Diehard tests on all bit streams \cite{sm}. The average pass rate over 10 separate tests with different data sets is 93\%, comparable to the pass rate of pseudo-RBG. 
 
To investigate the effect of photodetection noise on RBG, we define the signal-to-noise-ratio (SNR) as the bin size $s$ for digitization of $\Delta I_n$ divided by $2\sigma$ of the background fluctuation in a channel. With the SNR much higher than 1 for all spatial channels, the random bits are generated predominantly by the laser emission, and a numerical estimation of the noise contribution is presented in \cite{sm}. Figure~\ref{fig3}F shows that the percentage of all 188 subtests that every bit stream passes is uncorrelated with the SNR for the pair of spatial channels XOR'ed to create it, indicating that the level of detection noise does not affect the random bit quality.
 
To confirm the absence of correlations among the parallel bit streams, Fig.~\ref{fig3}D shows that the mutual information (MI) between any pair of bit streams \cite{sm} is as small as the MI of uncorrelated bit streams from different lasers. Moreover, to exclude short-term correlations, we combine odd bits from one stream and even bits from another to generate new sequences. The NIST tests of such combined bit sequences yield a pass rate of 72 to 73$\%$~\cite{sm}, demonstrating that all the original parallel bit streams are truly independent. 

All these test results certify the randomness of our parallel random bits generated at a cumulative rate of $2$~Tb/s\ $\times 127 = 254$~Tb/s. The very high RBG rate that we obtain indicates an enormous amount of entropy created by our laser. To establish its physical origin, we consider a simple model including only the interference of transverse and longitudinal lasing modes and spontaneous emission noise \cite{sm}. Using the Cohen-Procaccia algorithm \cite{hart2017recommendations}, we estimate the entropy rate $h_\mathrm{CP}$ for a bit stream generated from the simulated intensity fluctuations of a single spatial channel. Figure~\ref{fig4}A shows the convergence of $h_\mathrm{CP}$ for different embedding dimensions $d$. Both the interference of a large number of lasing modes and the spontaneous emission noise contribute to entropy generation \cite{sm}. As a result of stochastic intensity fluctuations, $h_\mathrm{CP}$ increases linearly with the number of digits $N_\mathrm{digit}$. The fact that $h_\mathrm{CP}$ reaches the information theoretical limit $h_0$ \cite{hart2017recommendations, sm} indicates that the maximal possible bit rate for a single channel is achieved. This rate exceeds the experimentally obtained value because of the limited temporal resolution and dynamic range of the photodetector.

To determine how many independent spatial channels are available for parallel RBG, we investigated the effective spatial degrees of freedom (DoFs) of the emission pattern of our laser. Intuitively, the number of spatial DoFs in the total intensity pattern is expected to be $2M$, where $M$ is the number of transverse lasing modes \cite{sm}. However, as a consequence of gain competition and saturation, the mode amplitudes are not uniformly distributed, effectively reducing the spatial DoFs. Applying the Karhunen-Loeve decomposition to the intensity pattern $I(x,t)$, we compute the Shannon entropy $H$ to obtain the complexity as a function of the number of transverse modes $M$ \cite{franz2007changing,sm}. In Fig.~\ref{fig4}B, the number of effective DoFs $2^H$ grows linearly with $M$, but with a slope smaller than 2. By maximizing $M$ with our cavity design, the maximal number of spatial channels is available for parallel RBG. Keeping only three LSBs after digitizing the emission intensity further reduces the spatial correlation length, and the number of independent channels is thus further increased \cite{sm}. 

In this proof-of-concept experiment, we have demonstrated parallel RBG in 127 independent channels with a rate of 2~Tb/s per channel. Both the single-channel bit rate and the number of spatial channels are limited by the resolution and efficiency of our experimental apparatus. Improving the temporal resolution and the dynamic range of photodetection can double the single-channel bit rate to $\sim$4~Tb/s. If all the emission is collected with finer spatial resolution, our laser can produce $\sim$500 independent bit streams \cite{sm}. Then the cumulative bit rate will reach 2~Pb/s.

It is possible to create a compact parallel-RBG system by integrating fast photodetectors with the laser in a single chip \cite{sm}. Alternatively, commercially available linear arrays of photodiodes may be butt-coupled to the laser chip on both ends. Although current photodiodes are not fast enough to fully resolve the temporal intensity dynamics, spatial multiplexing with hundreds of channels alone will drastically increase the RBG rate.

Compared to existing RBG schemes, our method, based on a single laser diode without optical feedback or optical injection, is extremely simple yet highly efficient. It does not necessitate any fine-tuning of operation parameters, and its performance is robust against fabrication defects. In our current experiments, the random bit streams are generated by a computer through off-line post-processing including XOR of bit streams from different locations. Real-time streaming of parallel random bits to a computer by conducting the post-processing (including XOR) ``on the fly'' remains a major technological challenge \cite{shinohara2017chaotic, ugajin2017real}. 

Besides the application of RBG, the extraordinary spatio-temporal complexity of our laser facilitates rich, diverse dynamical behavior, which can be finely tailored via the cavity geometry. By varying the spatial structure of cavity modes and tuning their characteristic length scale, we could effectively manipulate their nonlinear interactions with the gain medium to create deterministic spatio-temporal structures on demand. Such an ability to control the number of active modes and their nonlinear interactions promotes our laser as a model system to study many-body phenomena and for harvesting spatio-temporal quantum fluctuations. Because our laser possesses a variety of temporal and spatial scales, it may also be useful for studying optical turbulence with high Reynolds numbers. Despite having a high-dimensional phase space with a complex landscape, our laser is compact and may be used for reservoir computing and for creating physical unclonable functions (PUFs). \\

% References
% ---------------------------
%\bibliographystyle{Science}
%\bibliography{Refs}

\section*{Acknowledgments}
H.C.\ and K.K.\ thank R. Roy and M. Sciamanna for stimulating discussions. We acknowledge the computational resources provided by the Yale High Performance Computing Cluster (Yale HPC).

\textit{Funding.}
Supported by NSF grant ECCS-1953959; Office of Naval Research grant No. N00014-21-1-2026; National Research Foundation Competitive Research Program grants NRF-CRP-18-2017-02 and NRF-CRP-19-2017-01 and A*Star AME programmatic grant A18A7b0058 for the work at Nanyang Technological University; and Science Foundation Ireland grant 18/RP/6236. 

\textit{Author contributions.}
K.K. conducted the experiment, analyzed data, and prepared the manuscript; S.B. built the experimental setup and guided data analysis; Y.Z. and Q.J.W. did sample fabrication; S.G. and O.H. guided the analysis of semiconductor laser dynamics; and H.C. conceived the idea, initiated the project and supervised the research.

\textit{Competing interests.}
The authors declare no competing financial interests.

\textit{Data and materials availability.}
All data needed to evaluate the conclusions in the paper are present in the paper or the Supplementary Materials.

% --------------------------------------------
% Materials and Methods and Supplementary Text
% --------------------------------------------

\renewcommand{\theequation}{S\arabic{equation}}
\renewcommand{\thefigure}{S\arabic{figure}}
\renewcommand{\thetable}{S\arabic{table}}
\setcounter{figure}{0}
\setcounter{equation}{0}

\section*{Materials and methods}

\subsection*{Device fabrication}

We fabricate edge-emitting semiconductor lasers with a commercial GaAs/AlGaAs quantum well (QW) epiwafer (Q-Photonics QEWLD-808). The cavities with curved facets are defined by photolithography and etched by an inductively coupled plasma reactive ion etcher. The etch depth is 4 $\mu$m. The top metal contacts are fabricated by photolithography, Ti/Au deposition, and lift-off (see Ref.~\cite{kim2019electrically} for details). A scanning electron microscope (SEM) image of a fabricated device is presented in Fig.~S1B. 

The cavity length $L$ varies from 400 $\mu$m to 800 $\mu$m. The transverse width $W$ scales with $L$ as $W = L / \sqrt{2}$. The radius of the curved facets $R$ is also proportional to $L$, so that the cavity stability factor $g = 1 - L/R$ is fixed to $-0.74$. In Figs. 1E, 2, and 3, the cavity lengths are $L$ = 400~$\mu$m, 600~$\mu$m, and 800~$\mu$m, respectively. 

The wide-stripe lasers with planar facets ($g = 1$) in Fig.~1B are made by wafer cleaving. 

% Figure S01
\begin{figure}[t]
	\begin{center}
		\includegraphics[width = \linewidth]{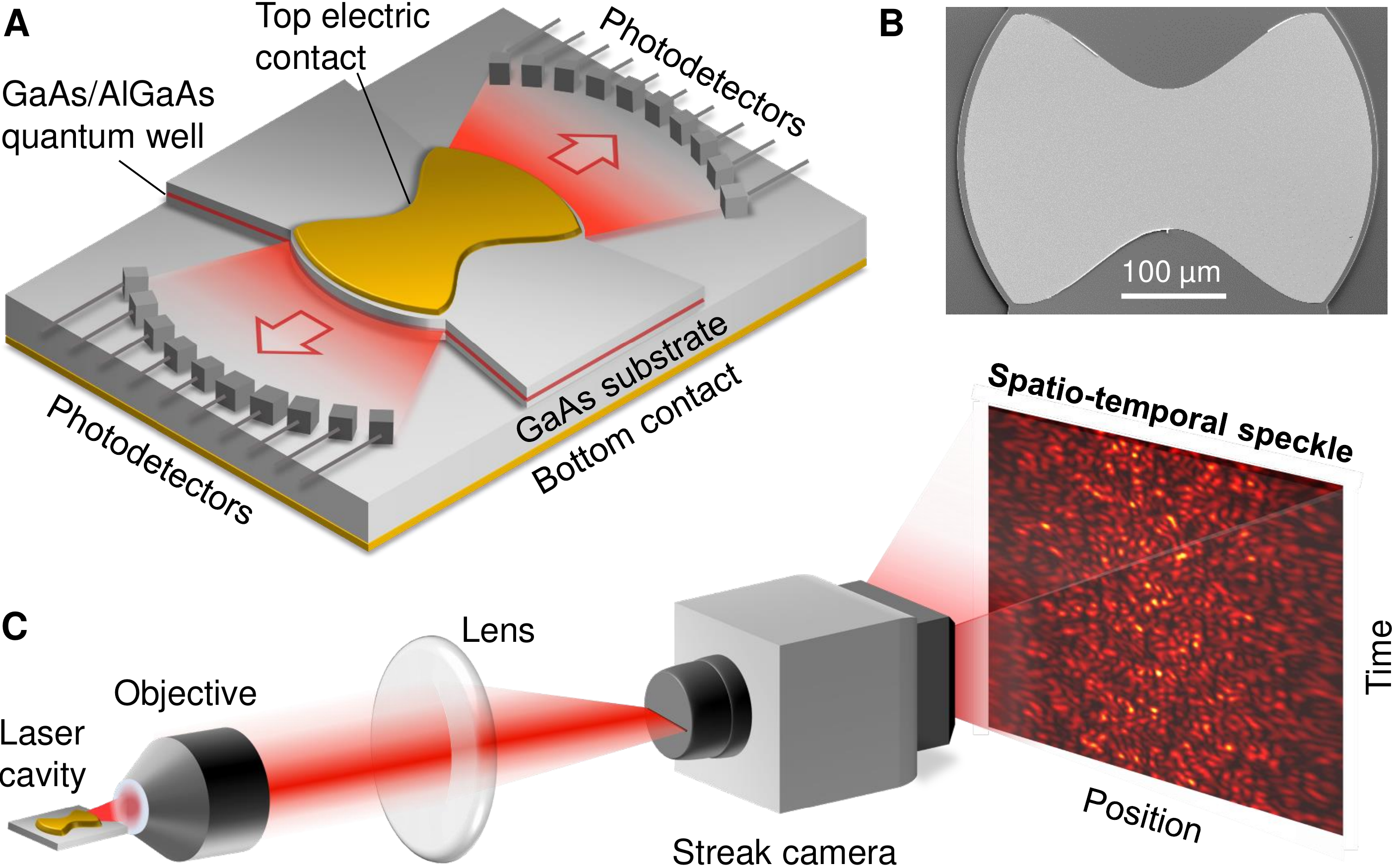}
	\end{center}
	\caption{Schematic of a chip-scale parallel random bit generator. 
		(\textbf{A})~Sketch of an electrically-pumped many-mode semiconductor laser with curved end facets. Its emission is measured by two arrays of photodetectors for parallel RBG. 
		(\textbf{B})~Top-view SEM image of a fabricated laser with cavity length $L$ = 400 $\mu$m, width $W$ = 282 $\mu$m, and end facet radius $R$ = 230 $\mu$m. 
		(\textbf{C})~Sketch of our experimental setup. The lasing emission at one end facet is imaged onto a streak camera that records the temporal fluctuation of the emission intensities at many spatial locations simultaneously.
	}
	\label{figS01}
\end{figure}

\subsection*{Device testing}

The diode laser is mounted on a copper plate. A tungsten needle (Quater Research H-20242) is placed on the top contact for electric current injection. To reduce sample heating, we use a diode driver (DEI Scientific, PCX-7401) to generate 150-ns-long current pulses at a repetition rate of 7 Hz. 

In Fig.~1B, the electric current injected to the stripe laser is 600 mA, where the lasing threshold current is 320~mA. In Figs.~1E, 2, and 3, the pump currents are set at two times the lasing thresholds, where the threshold currents are 400~mA, 600~mA, and 800~mA, respectively. 
 
As schematically shown in Fig.~\ref{figS01}C, the lasing emission on one end facet is imaged by a 20$\times$ microscope objective (NA = 0.4) and a lens (focal length 150 mm) onto the entrance slit of a streak camera (Hamamatsu C5680). The fast sweep unit (M5676) of the streak camera records the spatio-temporal traces of the emission intensity. Due to the limited field of view of the imaging optics, only the emission from the central part of the laser facet is collected. 

Because the facet is curved, the regions near the boundary of the field of view are out of focus when imaged onto the entrance slit of the streak camera. Thus the measured intensity of emission from these regions is lower than that from the central region of the facet. 

In the streak images shown in Figs.~1B\&E, the time window is 5-ns-long, and the temporal resolution is about 12~ps. Shortening the measurement time window to 0.5 ns in Fig.~2A increases the temporal resolution to 1.2 ps. 

Due to the finite length of the time windows measurable by the streak camera, we concatenate one million streak images from consecutive pulses to obtain a 0.5-ms-long time trace for each spatial position. This process does not artificially increase the entropy generation rate according to our numerical simulation detailed in the supplementary text. 

The emission spectrum is measured by an imaging spectrometer (Acton SP300i) equipped with an intensified CCD camera (Andor iStar DH312T-18U-73).

\subsection*{Spatio-temporal cross-correlation}
We calculate the correlation function $C(\Delta x,\Delta t)$ of the spatio-temporal emission intensity pattern $I(x,t)$,
\begin{equation}
C(\Delta x,\Delta t) = \frac{\left<\delta I(x_0 + \Delta x,t + \Delta t) \delta I (x_0, t)\right>_t}{\sqrt{\left<\delta I^2(x_0 + \Delta x,t)\right>_t\left<\delta I^2 (x_0,t)\right>_t}}   
\label{eqS1}
\end{equation}
where $\delta I(x,t) = I(x,t) - \left<I(x,t)\right>_t$ represents the temporal fluctuation of emission intensity at the transverse position $x$ on the end facet, and $x_0$ denotes the center of the facet~\cite{fischer1996complex}. In Figs.~1C\&F, $C(\Delta x,\Delta t)$ is averaged over 10 measurements.

\subsection*{Radio-frequency (RF) spectrum}
We calculate the RF spectrum of the emission intensity by performing the Fourier transform of a time trace at the spatial location $x = 0$ and taking the modulus. In Fig.~2C, the measured RF spectra of the lasing emission and background noise are averaged over 10 measurements, respectively. To be comparable to the experimental data, the simulated RF spectra are averaged over 10 realizations with different random phases (see the section Numerical modeling).

\subsection*{Random bit generation}

The lasing emission intensity in every spatial channel is temporally sampled at intervals of length $\tau$ (left inset of Fig.~3A). Faster temporal sampling will give a higher bit rate, but adjacent bits are correlated if the sampling period $\tau$ becomes shorter than the correlation time of the fluctuations. The optimal sampling period of $\tau$ is found to be 1.46 ps (see Fig.~\ref{figS08} and related discussion).

For spatial multiplexing, a smaller channel width $\Delta x$ yields a larger number $J$ of spatial channels. However, correlated bit streams will be produced if $\Delta x$ is shorter than the spatial correlation length of the lasing emission. A larger width also increases the signal in every channel. We find the optimal width $\Delta x$ = 1 $\mu$m (see Fig.~\ref{figS13} and related discussion). 

In a 2D streak image of the spatio-temporal intensity of the laser emission, each spatial channel of width $\Delta x$ = 1 $\mu$m contains 4 spatial pixels. The temporal sampling period $\tau$ = 1.46 ps corresponds to 3 temporal pixels. We sum the intensities of $4 \times 3 = 12$ pixels in a streak image to obtain one intensity value $I_n$. 

The streak camera background counts fluctuate on the spatial scale of a single pixel. The noise contribution to RBG is determined by its fluctuation, not its mean value which cancels out in the differential intensity. The standard deviation $\sigma$ of the differential background count (with the laser turned off) is 1.1. When summing over 12 pixels, the standard deviation is increased $\sqrt{12} = 3.46$ times to $\sigma$ = 3.9. 

Since $\sigma$ is much smaller than the bin size $s = 54$ of differential intensity $\Delta I_n$ in Fig.~3A, the random bits are determined predominately by the laser emission. However, if the value of $\Delta I_n$ is close to the boundary of one bin, the noise could alter the bit extraction. This is estimated to happen for $3.3 \%$ of the bits in the spatial channel shown in Fig.~3A. The percentage can be reduced by increasing the signal strength with better collection of the laser emission.

To further remove any residual bias in the parallel bit streams from 254 spatial channels, we divide the 254 spatial channels into two groups, and perform an XOR between the $j$-th spatial channel in the first group and the $(j+127)$-th spatial channel in the second group, where $j$ is an integer ranging from 1 to 127. Thus the total number of parallel bit streams is 127.

\subsection*{Mutual information} 
We calculate the mutual information between a pair of bit streams, $Y_i$ and $Y_j$, generated in parallel,
\begin{equation}
h_{ij} = \left<\! \sum_{Y_i, Y_j}p(Y_i,Y_j)\log_2\frac{p(Y_i,Y_j)}{p(Y_i)p(Y_j)} \!\right>.
\label{eqS2}
\end{equation}
Here $p(Y_i)$ is the probability density function (PDF) of a random bit stream $Y_i$, $p(Y_i, Y_j)$ is the joint PDF of two random bit streams $Y_i$ and $Y_j$, and $\left<...\right>$ denotes an average over all pairs of channels $i$ and $j$ with a constant spatial distance.

\subsection*{NIST tests of random bits} 

The NIST SP800-22 Random Bit Generator test suite consists of 15 different kinds of statistical tests, some of which include subtests~\cite{bassham2010sp}. Each test returns a single or multiple p-values. When the p-value exceeds a significance level of $\alpha$ = 0.01, the bit stream is considered random, as recommended by NIST. For $k$ bit streams, we examine if they pass or fail each statistical test. The pass proportion should be within $(1-\alpha) \pm 3\sqrt{\alpha(1-\alpha)/k}$. For each spatial channel, we use $k$ = 1000 bit sequences, each having $2^{20}$ bits, in total over $10^9$ bits. As an example, Fig.~\ref{figS15} shows the test results for one bit stream. In panel A, the pass proportions are all above the criterion indicated by the red line.

For a good random bit generator, the p-values from the $k$ bit streams should be uniformly distributed. The composite P-value (p-value of the p-values) is a measure of uniformity of the p-values. The distribution of p-values is considered as uniform when the composite P-value is larger than a significance level of $10^{-4}$. The example in Fig.~\ref{figS15}B shows that the composite P-values of all subtests are above the significance level indicated by the red line.

\subsection*{Diehard tests of random bits} 

The Diehard battery of randomness tests consists of 18 different kinds of statistical tests~\cite{marsaglia1996diehard}. Each test returns a single or multiple p-values. Some of the tests return a large number of p-values, and a composite p-value is calculated by the Kolmogorov-Smirnov (KS) test to determine if the p-values are uniformly distributed in [0,1). A bit sequence is considered random when p-values from all the tests are within the 95\% confidence interval of [0.0001, 0.9999]~\cite{tsoi2007high}. 
These tests require about 100 Mbit per channel, and we acquire sufficient data to perform the tests 10 times. The pass rates for all bit streams are listed in Table~\ref{tableS1}.

\subsection*{Numerical modeling} 

The cavity resonances are calculated with the eigenmode solver module of COMSOL Multiphysics. The passive modes are transverse-electric (TE) polarized, in accordance with the predominant polarization of GaAs quantum well lasers. To investigate the mode competition for optical gain, we calculate the lasing modes with the single pole approximation steady-state ab-initio lasing theory (SPA-SALT) \cite{ge2010steady, liew2015pump, cerjan2016controlling}. To maximize the number of transverse lasing modes, we optimize the cavity geometry: the ratio of cavity length to width $L/W = 1.41$, and the radius of the end facets $R/L = 0.58$~\cite{kim2019electrically}. The effective refractive index $n$ is set to 3.37. The total emitted field is a sum of fields in numerous transverse and longitudinal modes with frequencies within the GaAs QW gain spectrum. The intensity pattern at one facet can be written as  
\begin{equation}
I(x,t) = \left| \sum_{m=0}^{M-1}\sum_{q} A_{m,q} e^{i\phi_{m,q}(t)} \psi_{m}(x) e^{i2\pi\nu_{m,q}t} \right|^2,
\label{eqS3}
\end{equation}
where $\nu_{m,q}$ is the frequency of a mode with the transverse index $m$ and longitudinal index $q$, $\psi_{m}(x)$ represents its transverse field profile on the end facet, and $A_{m,q}$ and $\phi_{m, q}$ denote its global amplitude and phase. Due to spontaneous emission, $\phi_{m, q}(t)$ fluctuates in time, following a Wiener process,
\begin{equation}
\Delta\phi(t) = \phi(t+\Delta t) - \phi(t) = \sqrt{2 \pi \delta\nu \Delta t} \,  Z(t),
\label{eqS4}
\end{equation} 
where $\delta\nu$ is the lasing mode linewidth, $\Delta t$ is a discrete time step, and $Z(t)$ is a normal-distributed random number with a standard deviation of 1. For each mode, the initial phase $\phi(t=0)$ is randomly chosen in the interval [0, 2$\pi$) with a uniform probability density.

\subsection*{Entropy rate calculation}

Using the Cohen-Procaccia algorithm~\cite{cohen1985computing, hart2017recommendations}, we compute the entropy rate $h_\mathrm{CP}$ as a function of the bin size $\epsilon$ for intensity digitization and the temporal sampling period $\tau$. For a time trace of emission intensity $I(t)$ at a single position of the laser facet, we construct $d$-dimensional data sets by introducing time delays: $I_1 = I(t)$, $I_2 =I(t+\tau)$, $...$ , $I_d = I(t+(d-1)\tau)$.  Then we randomly select $N$ reference points in the $d$-dimensional space. For each reference point $j$, we compute $f_j(\epsilon)$, the fraction of other points within a $d$-dimensional box of width $\epsilon$. The $d$-dimensional pattern entropy estimate is given by 
\begin{equation}
H_d(\epsilon,\tau) = -\frac{1}{N}\sum^{N}_{j=1}\mathrm{log}_2[f_j(\epsilon)].
\label{eqS5}
\end{equation}
The Cohen-Procaccia entropy rate estimate is then obtained by 
\begin{equation}
h_\mathrm{CP}(\epsilon,\tau,d) = \tau^{-1}[H_d(\epsilon,\tau)-H_{d-1}(\epsilon,\tau)].
\label{eqS6}
\end{equation}
Here $\epsilon = (I_{\rm max}-I_{\rm min})/2^{N_\mathrm{digit}}$, with $I_{\rm max}$  and $I_{\rm min}$  being the maximum and minimum intensities, respectively.

In Fig.~4A, $h_\mathrm{CP}$ is plotted as a function of $N_\mathrm{digit}$ for different $d$. The time trace of emission intensity in a single channel is numerically calculated with the cavity parameters identical to the experimental ones ($L$ = 600 $\mu$m, $W$ = 424 $\mu$m, $R$ = 345 $\mu$m). There are 8 longitudinal mode groups within the emission spectrum. The number of transverse modes is $M$ = 200. The optical linewidth of each mode is $\delta\nu$ = 100 MHz. The sampling period is $\tau$ = 1.5 ps. The temporal range is 1.5 $\mu$s, yielding a time trace of $10^6$ samples. To have the same dynamic range as the experimental data, the intensity values are rounded to an integer number with a mean of 60, a minimum of 0, and a maximum of 300. The information theoretical limit $h_0$ is given by~\cite{gaspard1993noise, hart2017recommendations},
\begin{equation}
h_0 = \min(\tau^{-1}, 2f_{BW})\{N_{\mathrm{digit}}-D_{\mathrm{KL}}[p(I)||u(I)]\}
\label{eqS7}
\end{equation}
where $f_{BW}$ denotes the signal bandwidth, and $D_{\mathrm{KL}}[p(I)||u(I)]=\sum_{I}p(I)\log_2[p(I)/u(I)]$ is the Kullback-Leibler divergence~\cite{cover2012elements} between the intensity PDF $p(I)$ and the uniform PDF $u(I)$ within the same range of digitization.

\subsection*{Karhunen-Loeve decomposition} 

We perform a Karhunen-Loeve decomposition of the simulated lasing intensity pattern $I(x,t)$ in a cavity of length $L$ = 40 $\mu$m, width $W$ = 28.2 $\mu$m, radius of end facets $R$ = 23 $\mu$m and effective refractive index $n$ = 3.37. The mode frequencies $\nu_{m,q}$ and spatial field profiles $\psi_m(x)$ are obtained from the COMSOL calculation of cavity resonances. The amplitudes $A_{m,q}$ of individual lasing modes are calculated with SPA-SALT. Their phases $\phi_{m,q}$ are random numbers in the range of [0,2$\pi$). 

From the intensity fluctuation $\delta I(x,t) = I(x,t) - \langle I(x,t) \rangle_t$, the spatial covariance matrix $C_{ab}=\left<\delta I(x_a,t)\delta I(x_b,t)\right>_t$ is constructed and its eigenvalues $\lambda_{\alpha}$ are computed~\cite{hess1994spatio}. $\lambda_{\alpha}$ is sorted from high to low with the index $\alpha$, and it reflects the amplitude of the corresponding eigenmode in $I(x,t)$. The value of $\lambda_{\alpha}$ has a sudden drop at $\alpha = 2M$, where $M$ is the number of transverse lasing modes. Hence, the spatial degrees of freedom is $2M$, where the factor 2 stems from the independent degrees of freedom in the amplitude and phase of the field of one mode. 

Applying the Karhunen-Loeve decomposition to the simulated intensity pattern $I(x,t)$, we quantify the spatial complexity by the Shannon entropy of the eigenvalues~\cite{franz2007changing}
\begin{equation}
H = -\sum_{\alpha} {p_{\alpha}} \log_2{p_{\alpha}},
\label{eqS8}
\end{equation}
where $p_{\alpha} = \lambda_{\alpha}/(\sum_{\alpha}\lambda_{\alpha}$) is the normalized eigenvalue  (see Fig.~4B). The effective degree of freedom $2^H$ is less than $2M$ due to the different amplitudes of individual eigenmodes.

% ------------------
% Supplementary Text
% ------------------

\newpage
\section*{Supplementary Text}

\subsection{Experimental measurement}

\subsubsection{Device characterization}

\begin{figure}[b]
		\centering
		\includegraphics[width = \linewidth]{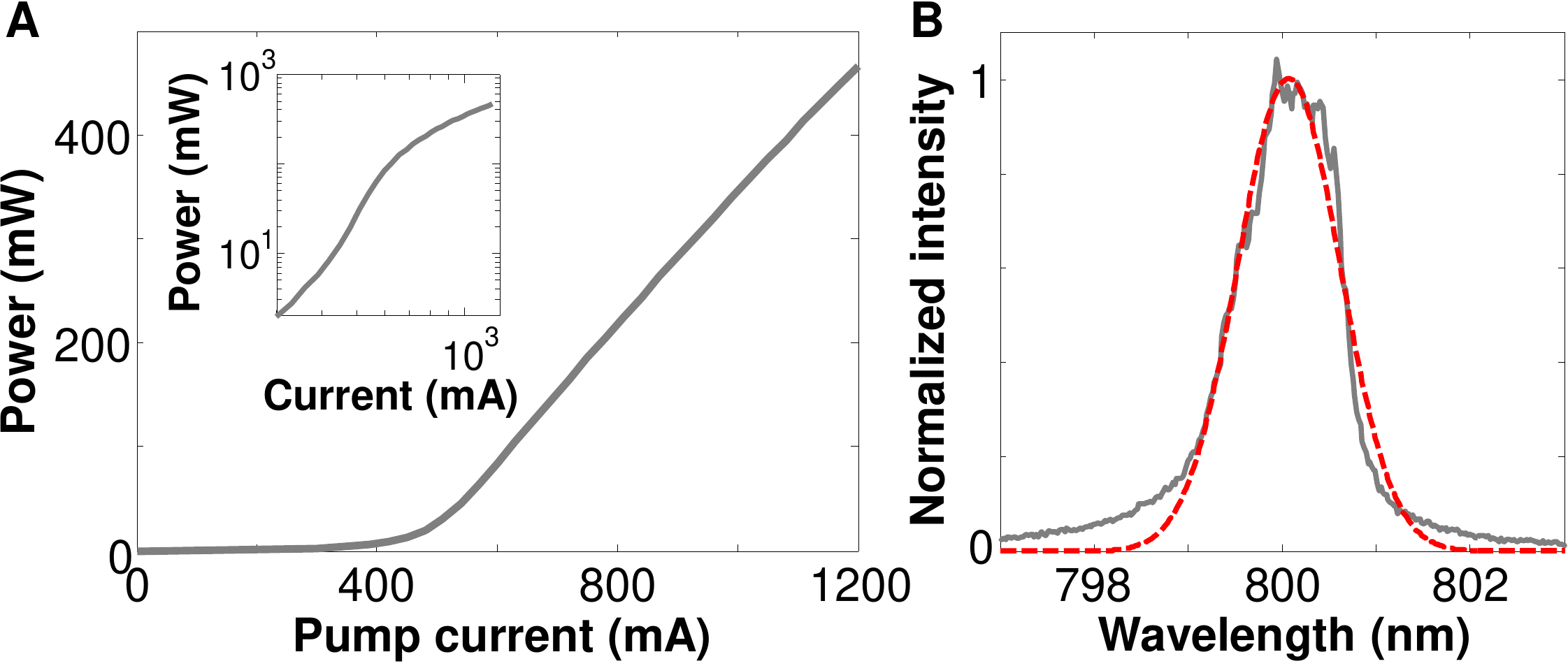}
		\caption{\textbf{Many-mode laser characteristics.}
			(\textbf{A})~LI curve: total emission power (from both facets of the laser) as a function of the electric current injected into the 600-$\mu$m-long cavity. The lasing threshold current is 570 mA. The inset is a double-logarithmic plot of the LI curve, exhibiting the S-shape that is a hallmark of lasers.
			(\textbf{B})~Time-integrated spectrum of lasing emission at a pump current of 1200 mA (gray solid line). The FWHM is 1.3 nm. The spectrum $S(\lambda)$ is fitted by a Gaussian function (red dashed line).} 
		\label{figS02}
	\end{figure}

We test lasing in the broad-area semiconductor lasers with curved facets. Similar LI curves and emission spectra are obtained for multiple lasers of identical geometry but varying size. Fig.~\ref{figS02}A shows the measured LI curve for a cavity of length $L$ = 600 $\mu$m, width $W$ = 424 $\mu$m, and radius of the curved facets $R$ = 345 $\mu$m. The lasing threshold current is 570 mA. At a pump current of 1200 mA, the power of lasing emission collected by the objective lens (NA = 0.4) from one end facet of the cavity is 93 mW. The collection efficiency is estimated to be about 20$\%$, with the divergence angle of the lasing emission, the transmission and numerical aperture of the objective lens, and the collection from a single facet of the cavity all taken into account. Thus the total emission power is about 470 mW, which corresponds to a quantum efficiency of 0.74 W/A. We record the far-field speckle pattern created by the output beam passing through a diffuser and calculate its intensity contrast~\cite{kim2019electrically}. The number of transverse lasing modes $M$ estimated from the measured speckle contrast of 0.07 is about 200. Hence the emission power per transverse mode is on the order of 1 mW.

    % FIGURE S02

Figure~\ref{figS02}B shows the measured emission spectrum at two times the lasing threshold integrated over a 0.6-$\mu$s-long pulse. Individual lasing modes cannot be resolved as the wavelength separation between adjacent modes is smaller than the resolution of the spectrometer. The full-width-at-half-maximum (FWHM) of the emission spectrum is 1.3 nm. The spectrum $S(\lambda)$ is fitted by a Gaussian function centered at $\lambda_0$ = 800 nm with a standard deviation of 0.55 nm. The free spectral range (longitudinal mode spacing) of the 600-$\mu$m-long cavity is $\lambda_\mathrm{FSR} = \lambda_0^2/2nL$ = 0.16 nm, where the effective refractive index $n = 3.37$ is calculated from optical confinement in the direction perpendicular to the cavity plane for transverse electric (TE) polarization, i.e., electric field parallel to the cavity plane. Hence the emission spectrum $S(\lambda)$ contains 8 longitudinal modal groups within its FWHM.

\subsubsection{Lasing dynamics}

In a conventional broad-area edge-emitting semiconductor laser, 
lasing occurs only in the lower order transverse modes because the higher order ones experience stronger diffraction losses and less gain. Furthermore, modulational instabilities induced by nonlinear interactions of the lasing modes with the gain material entail irregular pulsation and filamentation~\cite{ohtsubo2012semiconductor, fischer1996complex}. As shown in Fig.~1B, the emission from a wide-stripe GaAs quantum well (QW) laser is spatially concentrated at multiple locations, forming wire-like streaks called filaments. They originate from carrier depletion in a transverse region of high lasing intensity, which leads to a local increase in the refractive index. The ensuing lensing effect causes self-focusing of the optical field, which evolves into a filament with a transverse size of several microns. The filaments are inherently unstable, and their intensities oscillate irregularly on a sub-nanosecond time scale. Since the spatio-temporal scales of these self-organized structures are determined by intrinsic properties of the active gain medium~\cite{Hess1996semiconductor, marciante1997spatio}, they fundamentally limit the parallel RBG rate.

To achieve massively parallel ultrafast RBG, we greatly increase the number of transverse lasing modes and eliminate the long-range spatio-temporal correlations of the emission intensity by tailoring the cavity geometry. In a cavity of increased width and curved facets, high-order transverse modes are effectively confined. The spatial overlap of these modes with the injected carriers is increased by adapting the shape of the top metal contact. Fine tuning of the facet curvature maximizes the number of transverse lasing modes~\cite{kim2019electrically}. The higher order transverse modes have a transverse wavelength $\lambda_t \sim 1 \mu$m, which reduces the lateral width of spatial holes burnt in the carrier density. The resulting refractive index changes induce optical lenses that are too small to focus light and create filaments. In addition, the spatial modulations of the refractive index on such short scales supersede and disrupt the large lenses induced by lower order transverse modes and thus prevent filamentation~\cite{bittner2018suppressing}. Consequently, long-range spatio-temporal correlations disappear in $C(\Delta x,\Delta t)$ (Fig.~1F). \\

\subsection{Numerical modeling}

    % FIGURE S03
	\begin{figure*}[t]
		\centering
		\includegraphics[width = 16.0 cm]{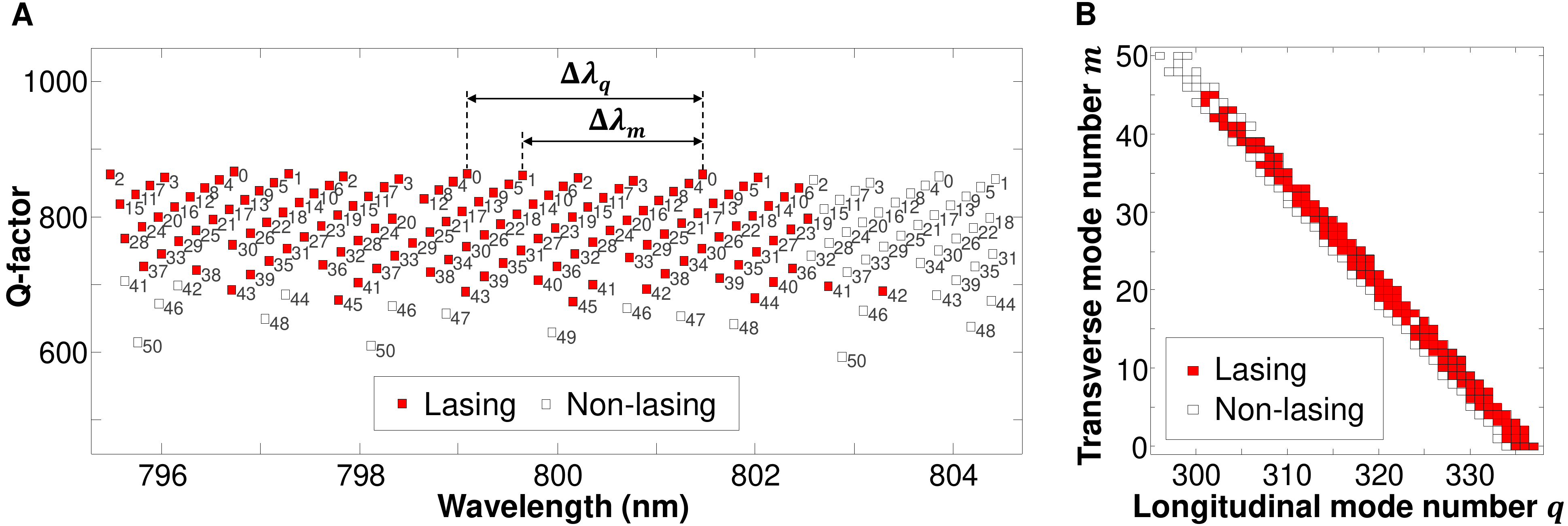}
		\caption{\textbf{Cavity resonances and lasing modes.}
			(\textbf{A})~Calculated $Q$ factors and wavelengths of the resonant modes (black open squares) in a cavity with length $L = 40~\mu$m, width $W = 28.2~\mu$m, curvature radius of the end facets $R = 23~\mu$m, and effective refractive index $n$ = 3.37. The transverse mode number $m$ is written next to each mode. The longitudinal mode spacing (free spectral range) $\Delta\lambda_{q}$ and the transverse mode spacing $\Delta\lambda_{m}$ are indicated by arrows. The modes that lase for pumping at two times the lasing threshold are marked by red solid squares.
			(\textbf{B})~The cavity resonances (black open squares) and lasing modes (red solid squares) shown in (A) are arranged in terms of their longitudinal and transverse mode numbers. Almost all transverse modes lase even in the presence of gain competition.}
		\label{figS03}
	\end{figure*}

\subsubsection{Cavity resonances}

We calculate the resonances of the passive cavity with the COMSOL eigenfrequency solver module. Since the laser emission is purely TE polarized, we compute only TE-polarized modes which are the solutions of the scalar Helmholtz equation
\begin{equation} 
    [\nabla^2 + k^2 n^2(x,y)] H_z(x,y) = 0 
    \label{eqS9} 
\end{equation}
with outgoing wave boundary conditions, where $k$ is the free-space wave number, $n$ is the effective refractive index, and $H_z$ is the z-component of the magnetic field. As light can escape in the transverse direction, absorptive boundary conditions are imposed on the lateral sides of the cavity using perfectly matched layers. Since numerical modeling of a cavity as large as the fabricated lasers ($L$ = 600 $\mu$m, $W$ = 424 $\mu$m) is too computationally expensive, we instead simulate a smaller cavity of the same geometry ($L$ = 40.0 $\mu$m, $W$ = 28.2 $\mu$m). The effective refractive index $n$ is equal to 3.37. We calculate the resonant modes with wavelengths around 800 nm which is similar to the measured laser wavelength (Fig.~\ref{figS02}B). However, since the smaller cavity has a larger mode spacing, we extend the wavelength range of simulation to 795$-$805 nm in order to increase the number of modes, covering 4 longitudinal mode groups. 
	
Figure~\ref{figS03}A shows the quality ($Q$) factors and wavelengths of the passive cavity modes. The transverse mode numbers are indicated. In Fig.~\ref{figS03}B, the modes are arranged in terms of their transverse and longitudinal mode indices. Although its size is relatively small, the simulated cavity exhibits about 50 transverse modes. The number of transverse modes scales linearly with the cavity size~\cite{kim2019electrically}.

\subsubsection{Lasing modes} 

Mode competition for optical gain tends to reduce the number of transverse lasing modes. We calculate the lasing modes using the single pole approximation steady-state \textit{ab-initio} laser theory (SPA-SALT) \cite{ge2010steady, liew2015pump, cerjan2016controlling}, taking gain saturation fully into account. We assume a spatially uniform distribution of the pump and a flat gain spectrum within the wavelength range of 795$-$805~nm. 

The modes that lase in the simulation are marked by red squares in Fig.~\ref{figS03}. When the pump is two times the lasing threshold, the number of transverse lasing modes $M$ is 46. Hence, most of the transverse modes in the cavity can lase in spite of gain competition.

\subsubsection{Many-mode interference} 

With spatio-temporal instabilities and filamentation suppressed in our laser, the lasing modes correspond to the high-$Q$ resonances of the passive cavity. To simulate a large cavity of size equal to the fabricated one ($L$ = 600 $\mu$m), we compute the mode frequencies $\nu_{m,q} = c/\lambda_{m,q}$ with the analytical expression for Hermite-Gaussian modes~\cite{siegman1986lasers},
\begin{equation}
\nu_{m,q} = \frac{c}{2nL} \left[q + \frac{1}{\pi} \left(m+\frac{1}{2} \right)\arccos(g)\right]
\label{eqS10}
\end{equation}
where $g \equiv 1 - L/R$ is the cavity stability parameter, which is $g = -$0.74 for the lasers considered here. 

The longitudinal mode spacing is $\Delta \nu_q = \nu_{m,q+1} - \nu_{m,q} = c/ 2nL$, and the transverse mode spacing is $\Delta \nu_m = \nu_{m+1,q} - \nu_{m,q} = (c/ 2nL) \arccos(g) / \pi$. Their ratio is $\Delta \nu_m/ \Delta \nu_q =  \arccos(g)/ \pi$. It is an irrational number for $g$ = $-0.74$, making the longitudinal mode spacing incommensurate with the transverse mode spacing. This result is confirmed by our numerical simulation of cavity resonances with COMSOL. We note that the frequency spacings are calculated for the modes of the passive cavity. Experimentally with electric current injection, the spatially inhomogeneous distribution of carriers can cause non-uniform changes of the refractive index, making the lasing mode frequencies deviate from the passive cavity mode frequencies. Nevertheless, the experimentally measured RF spectrum of the lasing emission in Fig.~2C remains broad and flat, similar to the numerically simulated spectrum of the passive cavity. Due to the presence of hundreds of lasing modes with many different frequency spacings between them, there is an almost continuous distribution of time scales in their interference pattern, so even if some of the time scales were by chance commensurate, the effect would be lost in the sea of other incommensurate ones. Therefore, we ignore such possible deviations in the following calculation.  

Summing over all lasing modes, the emission intensity at one spatial position, e.g., the center of the end facet $x_0$, is 
\begin{equation}
I(t) = \left| \sum_{m=0}^{M-1}\sum_{q} A_{m,q} e^{i [2\pi\nu_{m,q}t+ \phi_{m,q}(t)]} \right|^2
\label{eqS11}  
\end{equation}
with $A_{m,q}$ approximated by $\sqrt{S(\lambda_{m,q})}$, where $S(\lambda_{m,q})$ is the fit of the measured emission spectrum in Fig.~\ref{figS02}B. The total number of transverse modes is $M$ = 200 as in the experiments~\cite{kim2019electrically}.

\begin{figure}[t]
		\centering
		\includegraphics[width = \linewidth]{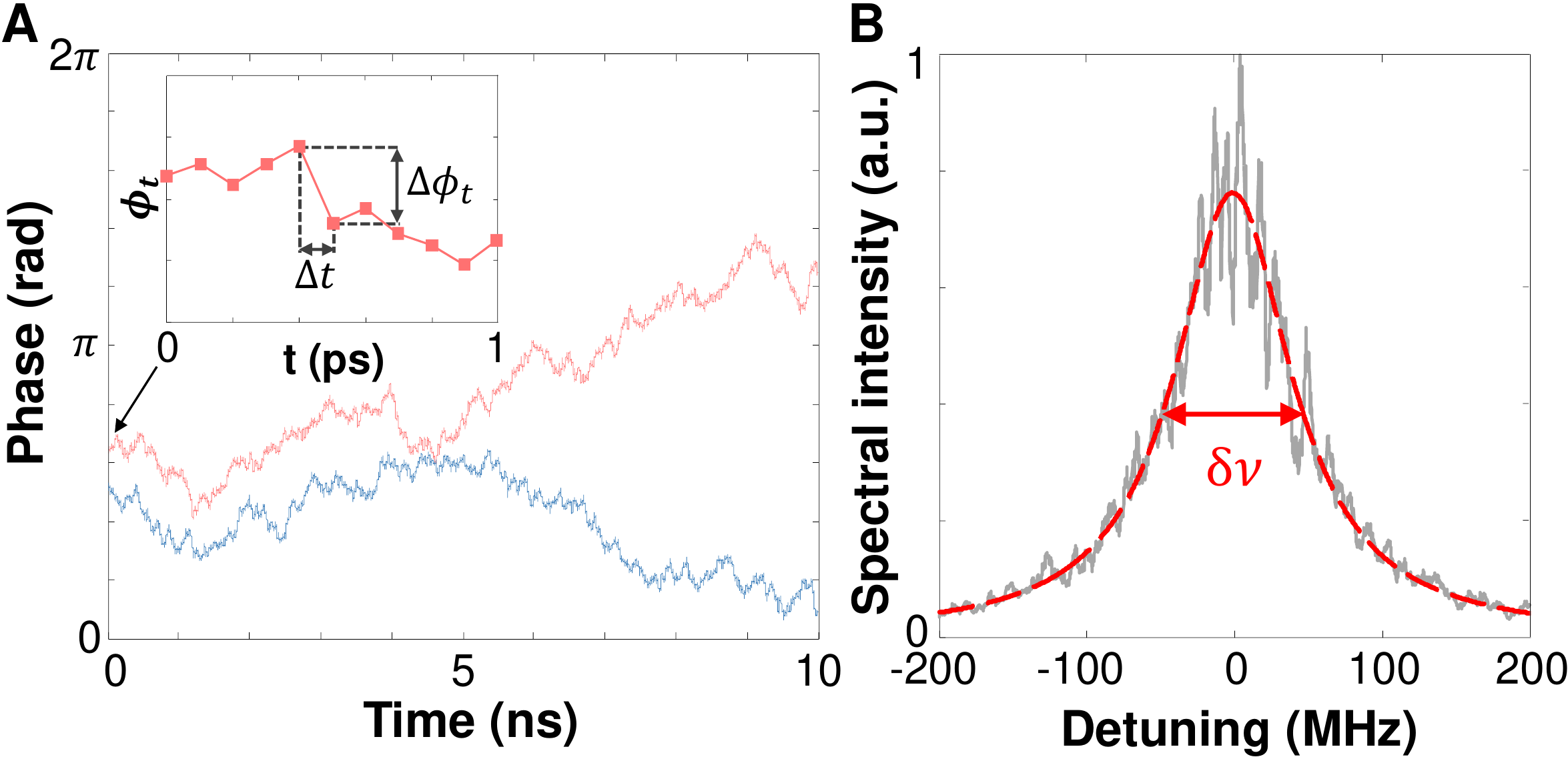}
		\caption{\textbf{Modeling of phase noise.}
			(\textbf{A})~Simulated random phase drift in time for two lasing modes. The inset shows the phase change $\Delta\phi_t$ after a single time step $\Delta t$.
			(\textbf{B})~The simulated optical spectrum of a lasing mode is fitted by a Lorentzian function with a FWHM of 96 MHz, which is close to the mode linewidth $\delta\nu = 100$ MHz.}
		\label{figS04}
	\end{figure}

To account for the spontaneous emission, we introduce a stochastic fluctuation to the phase $\phi_{m,q}(t)$ of each lasing mode~\cite{gallion1984quantum}. The mode linewidth is set to $\delta\nu$ = 100 MHz, which is typical for a GaAs/AlGaAs QW edge-emitting multi-mode laser~\cite{henry1986phase, elsasser1985multimode}. The corresponding coherence time, given by the inverse of the linewidth, is about 10 ns. We set the discrete time step $\Delta t$ to 0.1 ps. As shown in Fig.~\ref{figS04}A, the phase of each mode undergoes a random walk. The optical spectrum of a single lasing mode, calculated via the temporal Fourier transform of its field, exhibits a Lorentzian-shaped line with FWHM equal to $\delta\nu$ as shown in Fig.~\ref{figS04}B.

	% FIGURE S04

\subsubsection{Single-channel dynamics}	

\begin{figure}[t]
		\centering
		\includegraphics[width = \linewidth]{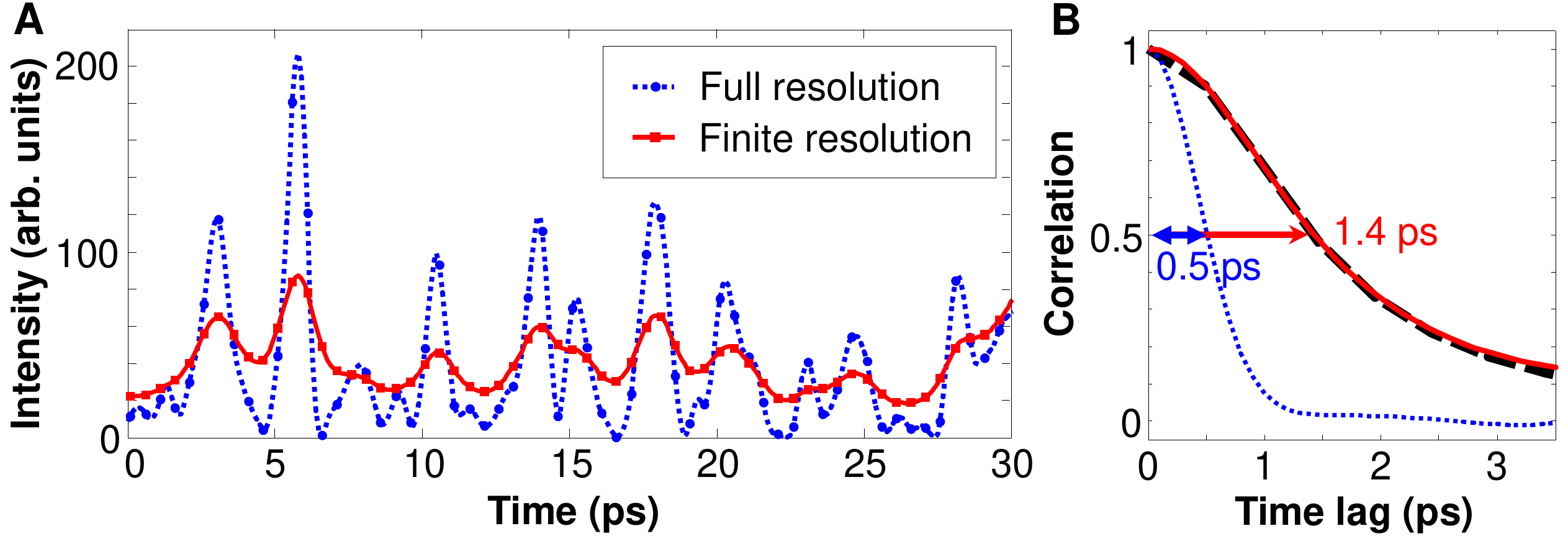}
		\caption{\textbf{Temporal fluctuations and correlation function.}
			(\textbf{A})~The time trace of the numerically simulated emission intensity in a single spatial channel (blue dotted line) is smoothed by convolution with the PSF of the streak camera (red solid line). The symbols represent sampling points separated by 0.5 ps.
			(\textbf{B})~Temporal correlation functions for the two traces in (A). The HWHM values are indicated. The black dashed line is the experimentally measured temporal correlation function.}
		\label{figS05}
	\end{figure}

Figure~\ref{figS05}A shows a portion of the simulated time trace $I(t)$ for a single spatial channel. The sampling period is set to 0.5 ps, corresponding to the temporal pixel size of our streak camera. We compute the temporal correlation function for $I(t)$ defined as 
\begin{equation}
C(\Delta t) = \frac{\langle \delta I(t) \delta I(t+\Delta t) \rangle_t}{\langle \delta I^2 (t) \rangle_t},
\label{eqS12}
\end{equation}
where $\delta I(t) = I(t) - \langle I(t) \rangle_t$ is the intensity fluctuation around the mean $\langle I(t) \rangle_t$. As shown in Fig.~\ref{figS05}B, the half-width-at-half-maximum (HWHM) of the correlation function is 0.5 ps. This is smaller than the measured correlation width (HWHM) of 1.4 ps due to the finite temporal resolution of the streak camera. 

To take into account the temporal resolution, we convolve the simulated time trace with a temporal point spread function (PSF). The PSF of the streak camera is approximated by a Lorentzian function with a FWHM of 1.2 ps. The convolution smoothens the time trace (Fig.~\ref{figS05}A), and increases the temporal correlation width to 1.4 ps, in agreement with the experimental value (Fig.~\ref{figS05}B). 

	% FIGURE S05

\subsubsection{Radio-frequency (RF) spectra}
	
To simulate the radio-frequency (RF) spectra shown in Fig.~2C, we calculate the Fourier transform of the simulated intensity $I(t)$ and take the modulus. The time trace is 500-ps-long, and the sampling period is 0.5~ps. The RF spectra are averaged over 10 time traces with different realizations of phase noise to be consistent with the measurements.

\begin{figure}[t]
		\centering
		\includegraphics[width = \linewidth]{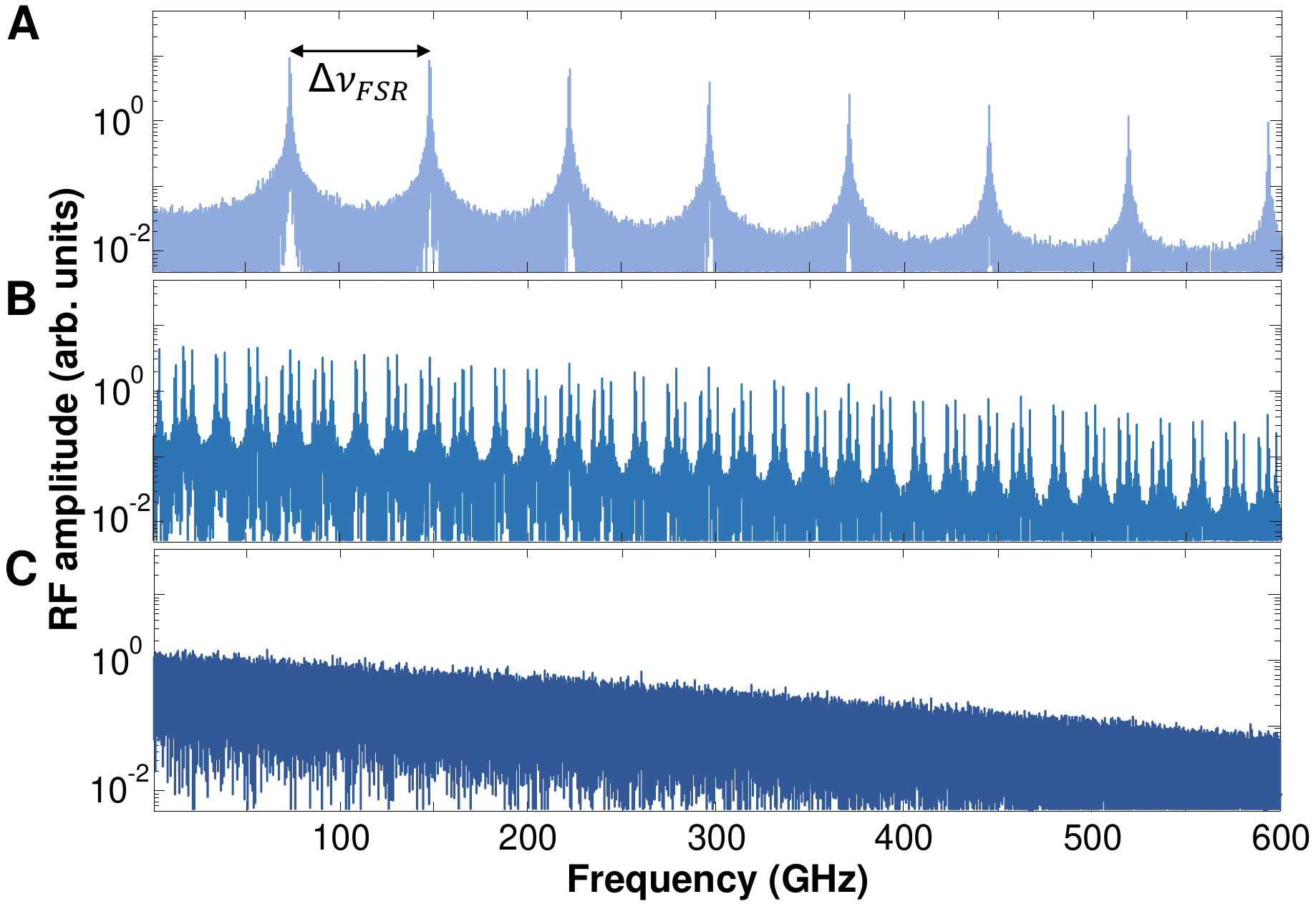}
		\caption{\textbf{Simulated RF spectrum.}
			 Fourier transform of the simulated emission intensity with the number of transverse modes (\textbf{A})~$M = 1$, (\textbf{B})~$M = 6$, and (\textbf{C})~$M = 200$. The number of longitudinal mode groups is fixed at 8. The spectrum becomes more and more densely packed and flat with increasing $M$. }
		\label{figS06}
	\end{figure}

To understand how the broad RF spectrum is formed, we vary the number of transverse modes $M$ while keeping the number of longitudinal mode groups constant. This corresponds to changing the width of a laser cavity while keeping its length fixed. With only the fundamental transverse mode $M$ = 1 (Fig.~\ref{figS06}A), the RF spectrum features multiple peaks separated by the longitudinal mode spacing (free spectral range) $\Delta\nu_{\mathrm{FSR}} = \Delta\nu_q = c/2nL$. With increasing $M$, each peak becomes a group of peaks that originate from the beating of transverse modes. For the cavity with $g$ = $-0.74$ considered here, the transverse mode spacing is incommensurate with the free spectral range because $\arccos(g)/\pi$ in Eq.~(\ref{eqS10}) is an irrational number. As a result, the additional beating frequencies that appear when increasing the number of transverse modes eventually fill the entire frequency range. With $M$ = 200 transverse modes in Fig.~\ref{figS06}C, the spectrum becomes continuous and featureless, in agreement with the experimental data in Fig.~2C. The dense packing of the RF spectrum increases the entropy generation as will be shown in the next subsection.  
	
	% FIGURE S06

\subsubsection{Entropy rate}

\begin{figure}[t]
		\centering
		\includegraphics[width = \linewidth]{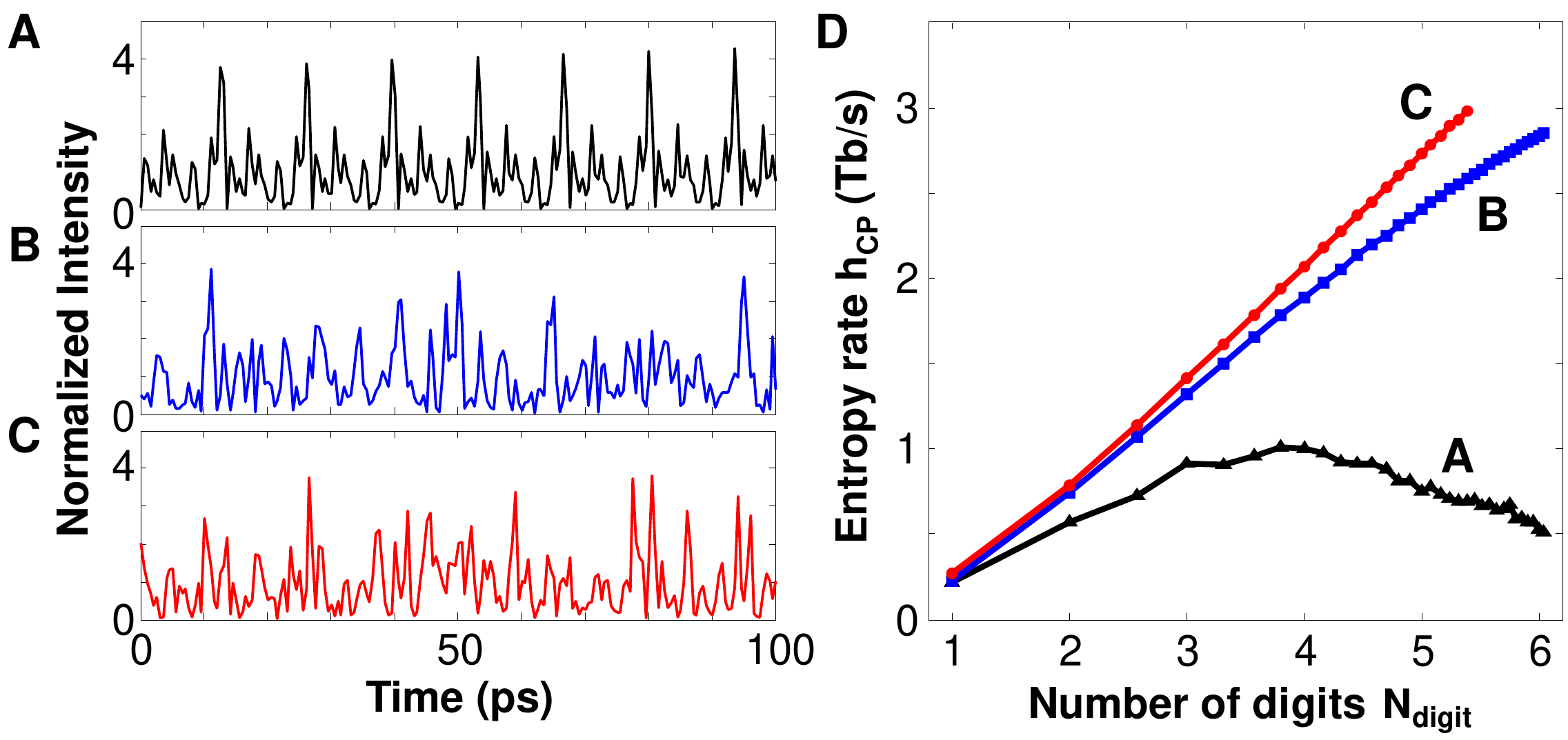}
		\caption{\textbf{Cohen-Procaccia entropy rate of simulated time traces.}
			(\textbf{A})~A portion of the simulated intensity time trace $I(t)$ with only the fundamental transverse modes ($M$ = 1), and without spontaneous emission noise.
			(\textbf{B})~$I(t)$ with $M$ = 3 and without spontaneous emission noise.
			(\textbf{C})~$I(t)$ with $M$ = 3 and with spontaneous emission noise ($\delta\nu$ = 100 MHz).
			(\textbf{D})~Entropy rate $h_\mathrm{CP}$ for the time traces in (A)-(C) with embedding dimension $d$ = 3. The sampling period $\tau$ is 1.5 ps.} 
		\label{figS07}
	\end{figure}

To understand how the number of transverse modes $M$ and the spontaneous emission noise affect entropy generation, we vary $M$ and turn on/off the phase noise when calculating the entropy rate. The number of longitudinal mode groups is fixed at 8 as before. We calculate the original entropy created by multimode interference using a time trace of intensity without accounting for the finite temporal measurement resolution. The length of the simulated time traces is 1.5 $\mu$s. With only one transverse mode and without spontaneous emission noise, the intensity trace is periodic in time (Fig.~\ref{figS07}A). The periodic modulation results from the temporal beating of the longitudinal modes with equal frequency spacing. Fig.~\ref{figS07}D shows the Cohen-Procaccia entropy rate $h_\mathrm{CP}$ versus the number of digits $N_\mathrm{digit}$ for embedding dimension $d$ = 3. As $N_\mathrm{digit}$ increases, $h_\mathrm{CP}$ first rises then drops, as the intensity trace repeats itself in time and no additional entropy is created eventually. With $M$ = 3 transverse modes, the intensity trace in Fig.~\ref{figS07}B exhibits more complex and aperiodic modulations. Since the transverse and longitudinal mode spacings are incommensurate, the temporal beating of 24 modes produces an intensity trace that will never repeat itself. Consequently, $h_\mathrm{CP}$ keeps increasing with $N_\mathrm{digit}$, first linearly then sublinearly in Fig.~\ref{figS07}D. Adding the spontaneous emission noise (Fig.~\ref{figS07}C) contributes to entropy generation as can be seen in a further increase of $h_\mathrm{CP}$ in Fig.~\ref{figS07}D. Moreover, $h_\mathrm{CP}$ grows linearly with $N_{\mathrm{digit}}$ as a result of the stochastic noise.\\

	% FIGURE S07
	
We note that the entropy rate is an average quantity obtained from a long time trace. Therefore, it cannot certify the randomness in any short time window, which is required to pass the NIST and Diehard tests. The details of randomness verification of our experimentally generated bit streams are presented in section 4.\\

\subsection{Random bit generation}

\subsubsection{Temporal sampling rate}

\begin{figure}[t]
		\centering
		\includegraphics[width = \linewidth]{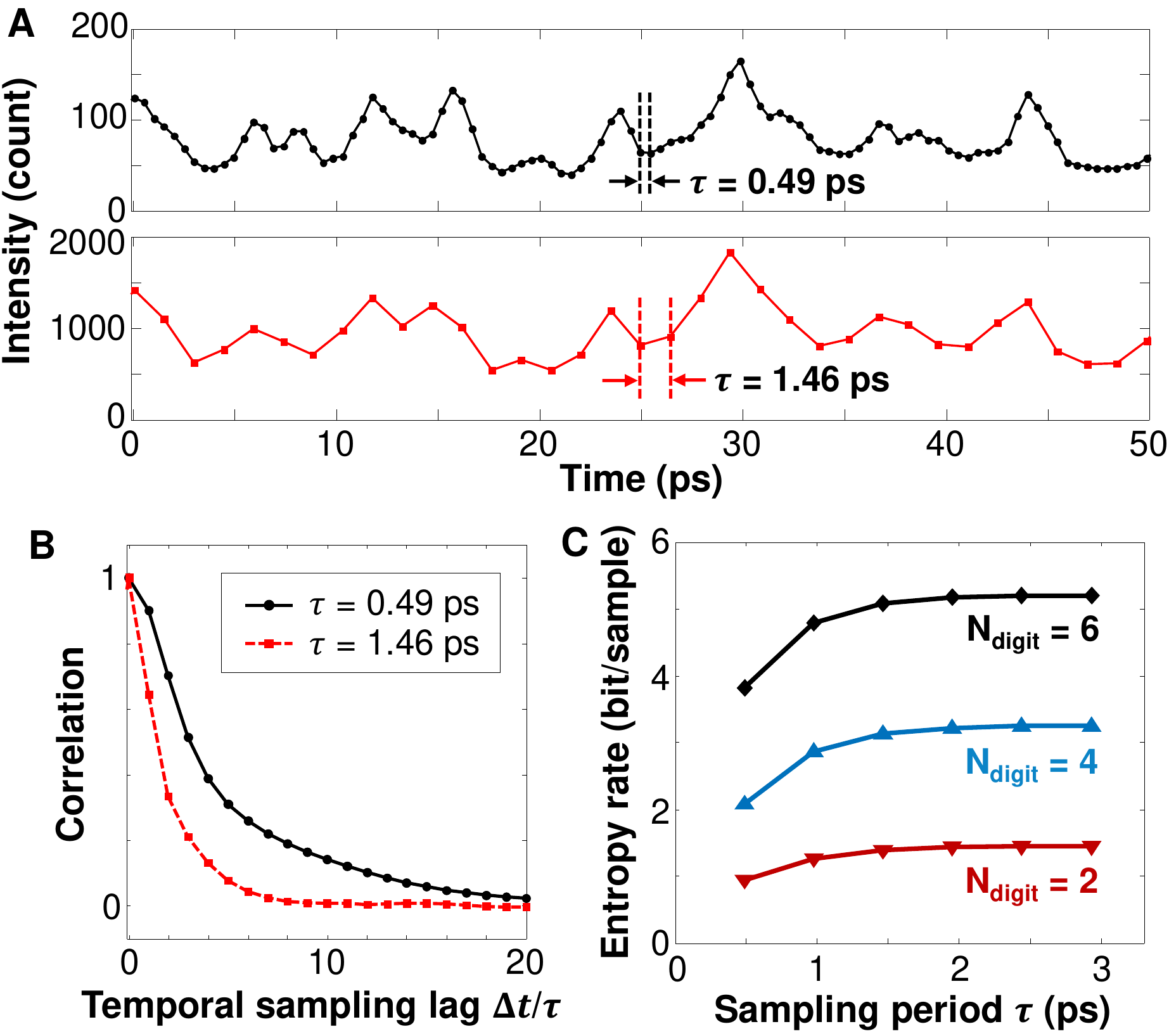}
		\caption{\textbf{Selection of temporal sampling period.}
			(\textbf{A})~Measured time trace of the emission intensity with a sampling period of $\tau$ = 0.49 ps (black), corresponding to a single pixel of the streak camera. To reduce the temporal correlation of sampled intensity, $\tau$ is increased to 1.46 ps (red). The intensity is integrated within the sampling period $\tau$ to increase the signal-to-noise ratio.
			(\textbf{B})~The temporal correlation functions for the two traces in (A) as a function of the time lag in units of the sampling period $\tau$ show a faster decay for the longer sampling period $\tau$.
			(\textbf{C})~Cohen-Procaccia entropy rate of a measured time trace for different sampling periods $\tau$ and numbers of digits $N_\mathrm{digit}$. The embedding dimension $d$ is set to 3.} 
		\label{figS08}
	\end{figure}

The black curve in Fig.~\ref{figS08}A is the emission intensity at one spatial position of the laser facet that is recorded by the streak camera. The temporal pixel size of the streak camera corresponds to 0.49 ps, and the sampling points are denoted by black dots. The temporal correlation function (Eq.~\ref{eqS12}) for this trace in Fig.~\ref{figS08}B reveals significant correlations among neighboring sampling points. To create independent bits, we must choose a longer sampling period $\tau$, which reduces the sampling rate and hence bit generation rate, however. 

	% FIGURE S08
	
In order to find the optimal sampling period, we calculate the Cohen-Procaccia entropy rate per sample for the experimental bit streams generated with different $\tau$. The sampled intensity  is binned into $2^{N_{\mathrm{digit}}}$ equally spaced intervals to create $N_{\mathrm{digit}}$ bits. As shown in Fig.~\ref{figS08}C, the entropy rate per sample first increases with the sampling period $\tau$, then levels off for $\tau \geq$ 1.46 ps. This trend is similar for different $N_{\mathrm{digit}}$. Hence we set the sampling period to be 3 times the streak camera pixel size, $\tau$ = 1.46 ps, to extract the maximal entropy per sample with the fastest possible sampling rate.
		
\subsubsection{Intensity PDF}
		
	% FIGURE S09
	\begin{figure}[b]
		\centering
		\includegraphics[width = \linewidth]{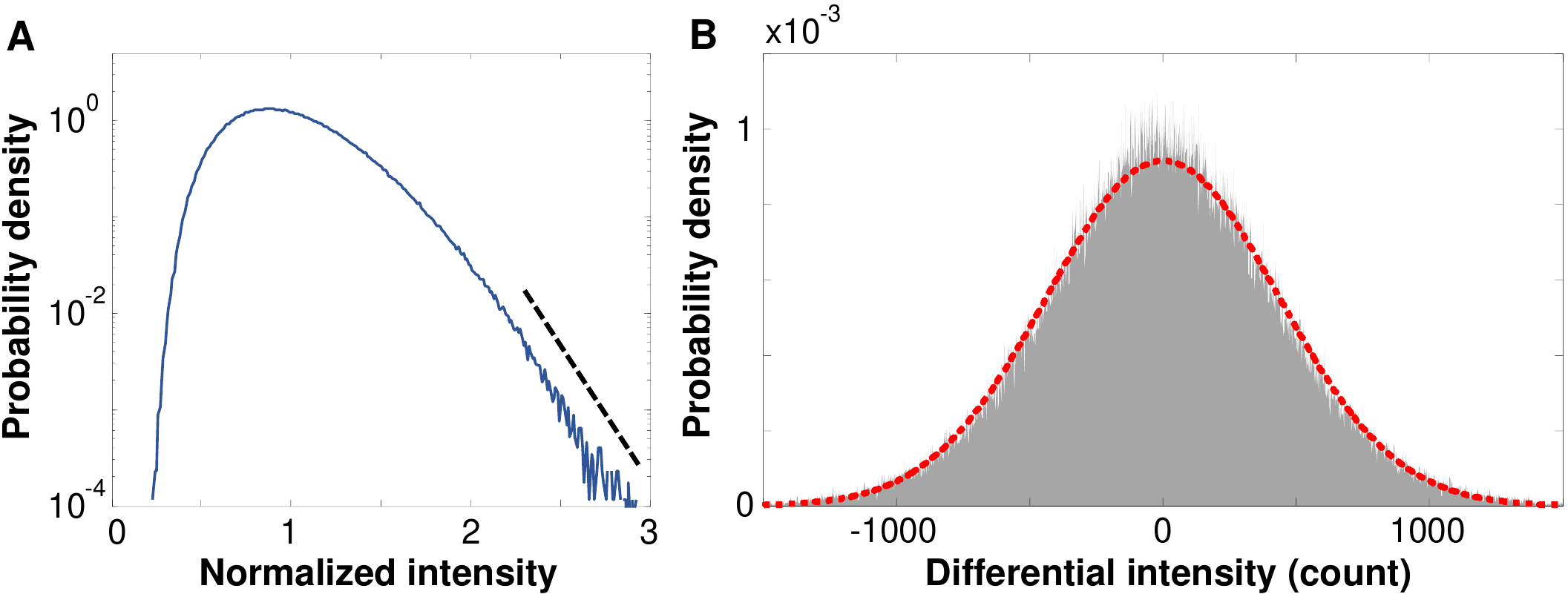}
		\caption{\textbf{Intensity PDF.}
			(\textbf{A})~The PDF of the measured intensity $I_n$ of emission in a single spatial channel. The temporal resolution is 1.46~ps and the spatial resolution is 1~$\mu$m. The black dashed line indicates an exponential decay.
			(\textbf{B})~The PDF of the differential intensity $\Delta I_n = I_{n+4} - I_n$ is symmetric and well fitted by a Gaussian function (red dotted line).} 
		\label{figS09}
	\end{figure}

Figure~\ref{figS09}A shows the probability density function (PDF) of the sampled intensity $I_n$ in one spatial channel. The PDF exhibits an exponentially decaying tail, which is a characteristic of Rayleigh speckle patterns. The PDF is not a perfect exponential function due to the finite spatial and temporal resolution of the measurement. This asymmetric PDF can lead to biased bits, which degrade the quality of random bits generated.

To make the distribution symmetric, we perform a subtraction of sampled intensities. We obtain a sequence of differential intensities $\Delta I_n = I_{n+m}-I_n$ by subtracting the intensities separated by a sample distance of $m$. For a small $m$, the PDF of $\Delta I_n$ deviates notably from a Gaussian function, because of the temporal correlation of $I_n$. The non-Gaussian PDF will introduce bias among different combinations of the three LSBs taken for RBG. A large $m$ produces a bit stream with long-range correlations, which also degrade the random bit quality. We choose $m$ = 4 as an optimal sample distance. Figure~\ref{figS09}B shows the PDF of the differential intensity from experimental data. It is fit well by a Gaussian function, which leads to the equal probability of 8 possibilities for 3 LSBs in Fig.~3B.

\subsubsection{Suppressing temporal correlations}

\begin{figure}[b]
		\centering
		\includegraphics[width = \linewidth]{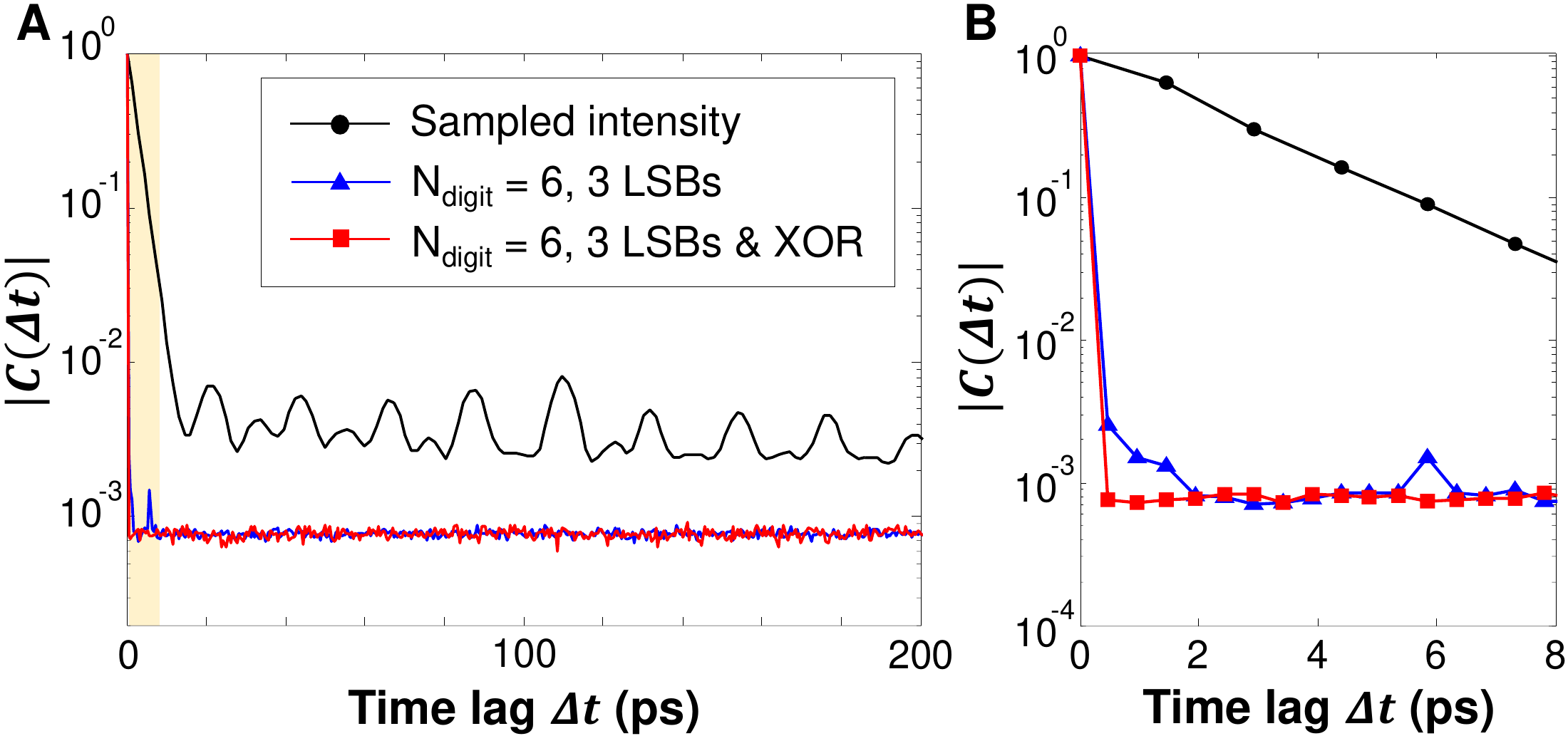}
		\caption{\textbf{Temporal correlation of bits in a single channel.}
			(\textbf{A})~The magnitude of the temporal correlation function $|C(\Delta t)|$ of the measured intensity trace sampled with $\tau$ = 1.46 ps (black circles), after keeping only the 3 LSBs (blue triangles), and after performing the XOR operation in addition (red squares). The correlation functions are averaged over spatial channels. The number of data points in time is $2^{20}$.
			(\textbf{B})~The magnified view for short time lags shows the quick decay of correlations for the three LSBs. The small peak at 5.84 ps (= 4$\tau$) is attributed to the subtraction of the sampled intensity $\Delta I_n = I_{n+4} - I_n$.
			}
		\label{figS10}
	\end{figure}

As seen in Fig.~\ref{figS08}B, even after choosing a sampling period of $\tau$ = 1.46 ps, the correlation of adjacent samples is not completely eliminated. RBG demands negligible correlation between successive bits. Here digitization and post-processing play a crucial role in removing the remaining correlation. During the analog-to-digital conversion (ADC), the measured intensity is transformed to $N_{\mathrm{digit}}$ = 6 bits, and only the three least significant bits (LSBs) are kept. In Fig.~\ref{figS10}, we compare the temporal correlation function [Eq.~(\ref{eqS12})] of the bit stream to that of the original intensity trace, averaged over all channels. For long time lags, keeping only the LSBs reduces the correlation below $10^{-3}$, which is the lower limit given by the finite length of the bit stream ($2^{20}$ samples). For short time lags (Fig.~\ref{figS10}B), the correlation for the LSBs decays rapidly with the sample distance, greatly shortening the correlation time. The residual correlations are then completely removed by the XOR operation with another bit stream from a distant spatial channel. The correlation remains at the background level for any time delay, indicating the absence of correlations between successive bits. 

	% FIGURE S10

\subsubsection{Number of spatial channels}
	
	% FIGURE S11
	\begin{figure}[b]
		\centering
		\includegraphics[width = \linewidth]{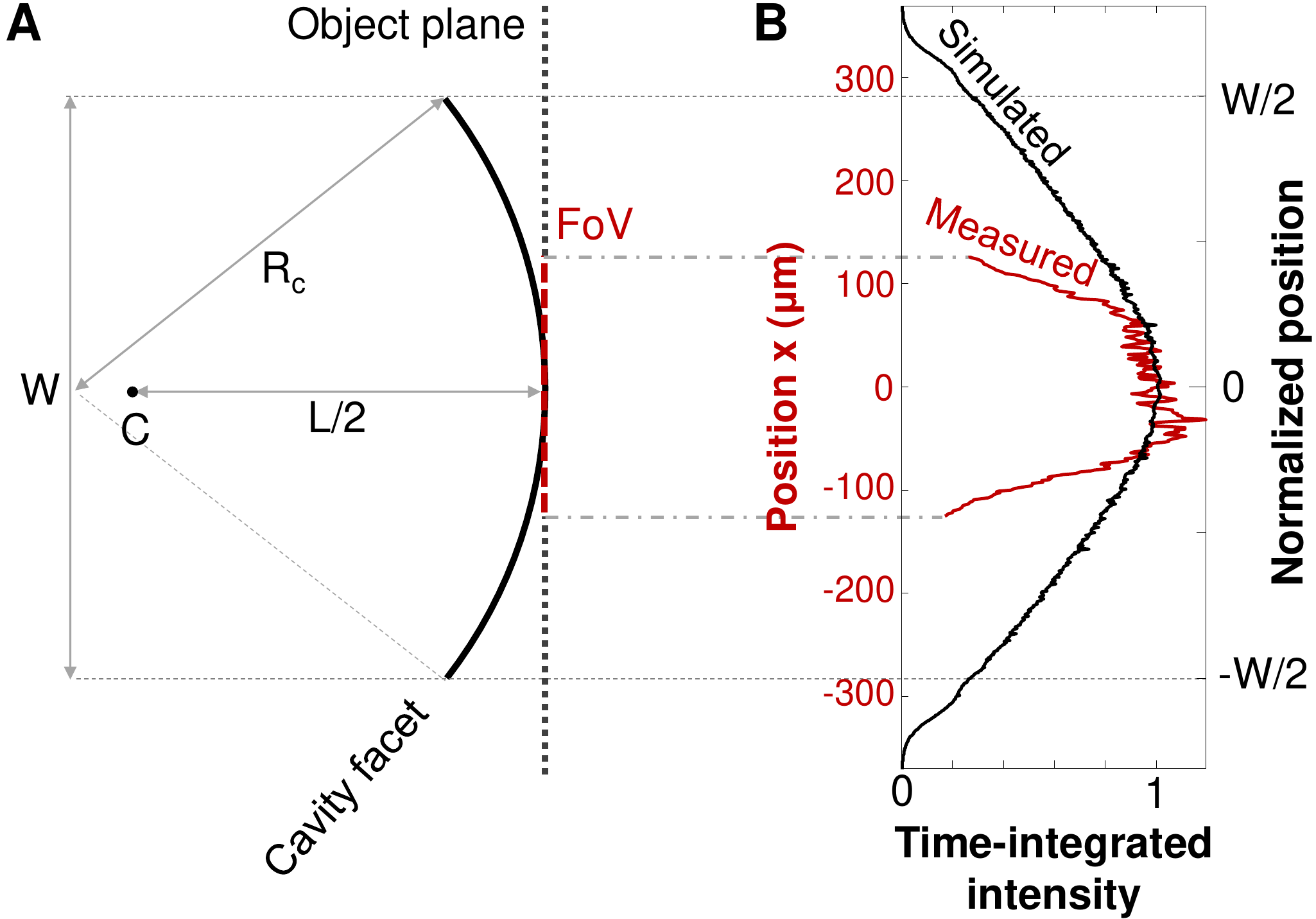}
		\caption{\textbf{Collection of spatial channels.}
			(\textbf{A})~The curved facet of the laser cavity is imaged onto the streak camera entrance slit (not shown). The two edges of the facet are away from the object plane (black dotted line). The field of view (FoV) indicated by the red dashed line is about half of the facet width.
			(\textbf{B})~The measured transverse profile of the emission intensity on the curved facet (red) is compared to the ideal profile from numerical simulations (black). Each curve is normalized by the intensity at $x = 0$.}
		\label{figS11}
	\end{figure}
		
Not all spatial channels are accessible in the current experiment because of the limited field of view (FoV) of the imaging optics. As shown in Fig.~\ref{figS11}A, the FoV is about 254 $\mu$m (marked by the red dashed line), which is less than half of the laser facet width $W$ = 566 $\mu$m. For RBG, we use all spatial channels within the FoV, which is less than half of the total channels available across one facet of the laser.
As the central part of the facet is imaged onto the streak camera, the curved parts close to the edges of the FoV are away from the object plane (marked by the black dotted line in Fig.~\ref{figS11}A) and are thus out of focus. Consequently, the emission intensity drops rapidly near the two edges of the FoV (red curve in Fig.~\ref{figS11}B).

In Fig.~\ref{figS11}B, we compare the measured emission profile to the ideal one obtained from numerical simulation. Using SPA-SALT, we calculate the lasing modes in two small cavities of lengths $L$ = 20, 40 $\mu$m with the same geometry as the laser cavities in the experiment. After rescaling, the calculated intensity distribution on the curved facet is identical for the two cavities. Hence, the emission profile is universal and scales linearly with the cavity size. The FWHM of the measured profile is about half of the simulated one. Hence, only half of the available spatial channels are used for random bit generation in the current experiment.

\subsubsection{Spatial correlation length}

	% FIGURE S12
	\begin{figure}[b]
		\centering
		\includegraphics[width = \linewidth]{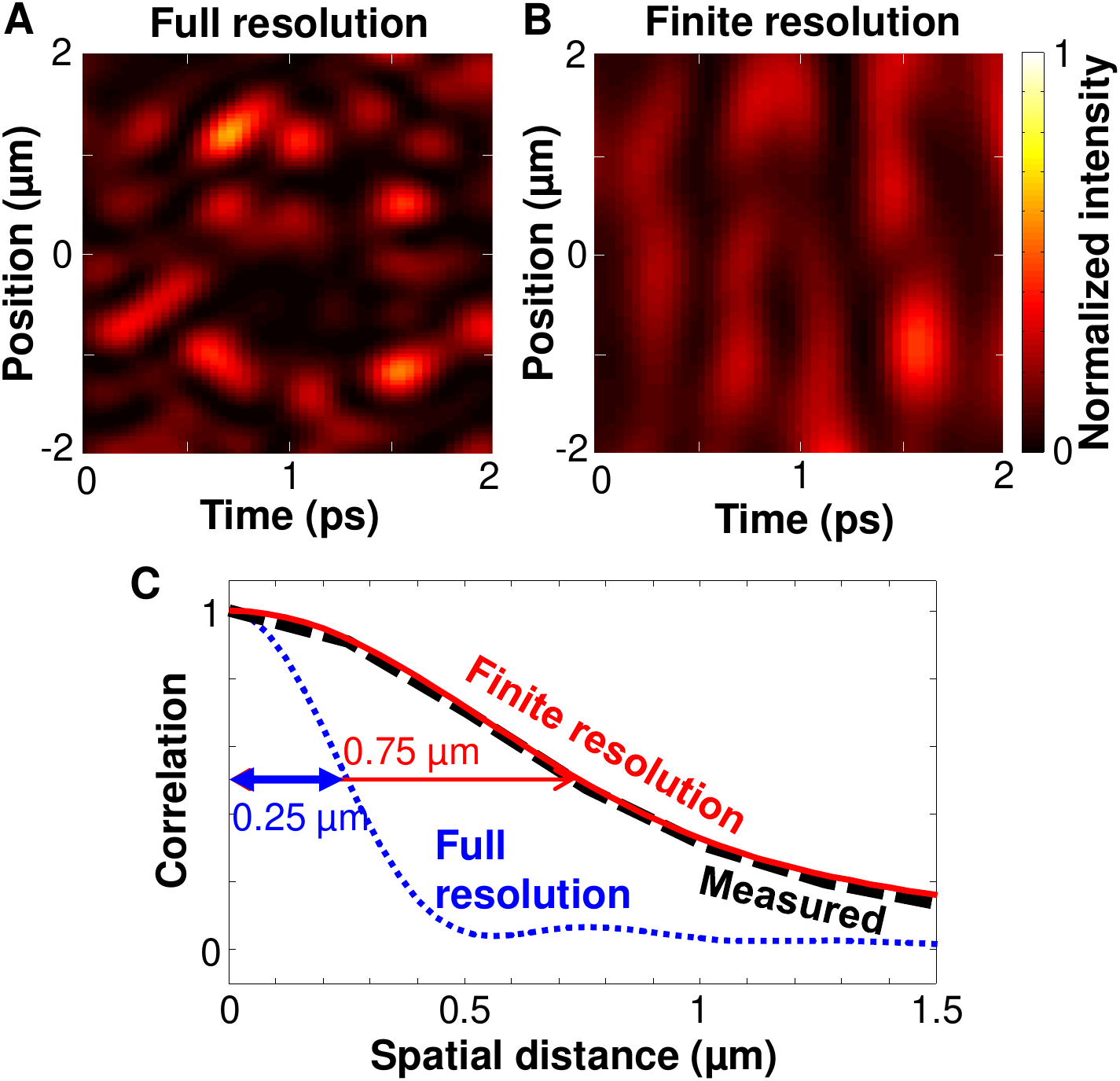}
		\caption{\textbf{Spatial correlation of simulated intensity fluctuations.}
			(\textbf{A})~The fully resolved spatio-temporal pattern of emission intensity on the end facet of the laser cavity features fine speckles. 
			(\textbf{B})~The intensity pattern after convolution of (A) with the spatial PSFs of the imaging optics (NA = 0.4) and the streak camera (Lorentzian with FWHM = 0.47 $\mu$m) shows an increase of the spatial speckle size.
			(\textbf{C})~The spatial correlation functions $C(\Delta x)$ for the intensity fluctuations in (A) (blue dotted line) and (B) (red solid line) have a HWHM of 0.25 $\mu$m and 0.75 $\mu$m, respectively. The black dashed line denotes the measured spatial correlation.}
		\label{figS12}
	\end{figure}
	
The spacing between independent spatial channels is determined by the spatial correlation length of the emission intensity. To find its relation to the transverse wavelength of the lasing modes, we simulate the spatio-temporal intensity pattern of the lasing emission from a small cavity with $L$ = 40 $\mu$m, $W$ = 28.2 $\mu$m and $R$ = 23 $\mu$m (see materials and methods). The number of transverse lasing modes $M$ at a pump level of two times the lasing threshold is 46 (see Fig.~\ref{figS03}). Figure~\ref{figS12}A is a portion of the spatio-temporal intensity pattern at one cavity facet with full spatial resolution. In the experiment, the numerical aperture (NA) of the imaging optics is 0.4, and the spatial resolution of the streak camera is about 0.5 $\mu$m. The convolution with the spatial point spread functions (PSF) of imaging optics and streak camera enlarges the spatial speckle grains (Fig.~\ref{figS12}B). In Fig.~\ref{figS12}C, we show the correlation of the intensity fluctuations $\delta I(x,t) = I(x,t) - \langle I(x,t) \rangle_t$ at spatial locations separated by $\Delta x$,
\begin{equation}
C(\Delta x) = \left<\! \frac{ \langle \delta I(x,t) \delta I(x+\Delta x,t) \rangle_t }{\sqrt{ \langle \delta I^2(x,t) \rangle_t \langle \delta I^2(x+\Delta x,t) \rangle_t }} \!\right>_{\!\!\!x} \, .
\label{eqS13}
\end{equation}

The HWHM of $C(\Delta x)$ is 0.25 $\mu$m for the fully resolved pattern in Fig.~\ref{figS12}A. The average speckle grain size (FWHM) is 0.5 $\mu$m, which is close to half of the transverse wavelength of the highest-order transverse lasing mode. The limited spatial resolution of imaging optics and streak camera enlarges the speckle grain size to 1.5 $\mu$m, thus reducing the number of independent channels.

\subsubsection{Suppressing spatial correlations}

\begin{figure}[b]
		\centering
		\includegraphics[width = \linewidth]{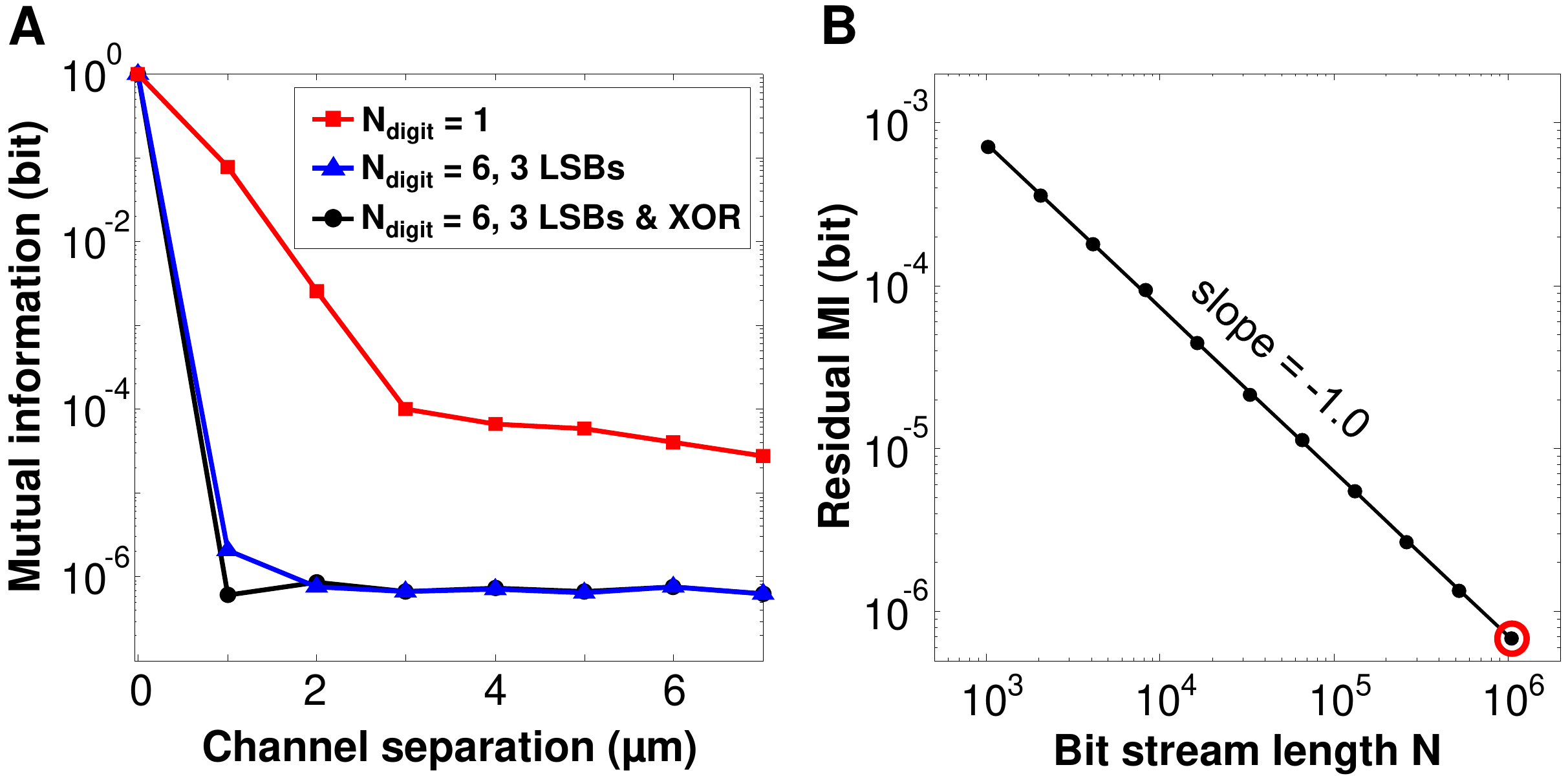}
		\caption{\textbf{Mutual information between two spatial channels.}
			(\textbf{A})~Mutual information (MI) of experimental bit streams produced by (i) thresholding (red squares), (ii) keeping 3 LSBs from $N_{\mathrm{digit}}$ = 6 bit conversion (blue triangles), (iii) calculating XOR of (ii) with a bit stream from a spatially distant channel (black circles). The MI is averaged over all channels. The MI decreases with the channel separation.
			(\textbf{B})~The residual mutual information of (iii) in (A) at channel separation $\Delta x \geq 1$ $\mu$m is inversely proportional to the bit stream length $N$, indicating it results from the finite stream length. The red circle indicates the length of the bit stream used in (A).}
		\label{figS13}
	\end{figure}

Digitization and post-processing reduce the spatial correlation length, similarly to the suppression of temporal correlation in Fig.~\ref{figS10}. In Fig.~3D, the mutual information of bit streams from two spatial channels becomes negligible when their separation $\Delta x$ exceeds 1 $\mu$m. However, the original intensity pattern in Fig.~2 shows a spatial correlation extending over a distance of 1.5 $\mu$m. In Fig.~\ref{figS13} we compare the mutual information (MI) between two bit streams produced experimentally for three different cases: (i) thresholding $N_{\mathrm{digit}}$ = 1, the simplest bit-extraction scheme; (ii) keeping 3 LSBs from analog-to-bit conversion with $N_{\mathrm{digit}}$ = 6; and (iii) conducting XOR of (ii) with a bit stream from a spatially distant channel. In comparison to (i), the MI between neighboring channels (with 1 $\mu$m spacing) is reduced by five orders of magnitude in (ii). Moreover, the MI is further reduced at short-range in (iii). It stays at the residual level of $10^{-6}$ and becomes independent of the channel separation. Fig.~\ref{figS13}B shows the residual MI is inversely proportional to the length of the bit stream $N$. For $N = 2^{20}$, the residual MI is less than $10^{-6}$ (circled in red). It is equal to the MI between any pair of channels in (iii), indicating all channels are statistically independent and their residual MI is a result of the finite bit stream length.

	% FIGURE S13

\subsubsection{Effect of noise on binning}

For every spatial channel, the entire binning range of its differential intensity $\Delta I_n$, which is set to 8.2 times the standard deviation of $\Delta I_n$, is divided into $2^6 = 64$ equally sized bins. If the value of $\Delta I_n$ is close to the boundary of one bin, the detection noise can alter the bit extraction. To examine how strong this effect is, we simulate the streak camera noise and add it to an experimental intensity trace. We compare the bit stream generated by the noise-altered trace to the original one. With 3 LSBs extracted from 6 digits, 3.3\% of all the bits in the spatial channel shown in Fig.~3A are altered by the addition of noise. For all 254 spatial channels, the average bit error rate is $4.1 \pm 1.2$\%. This percentage can be reduced by increasing the signal strength with better collection of the laser emission.

\subsubsection{Concatenating time traces}
	
    % FIGURE S14
	\begin{figure}[b]
		\centering
		\includegraphics[width = \linewidth]{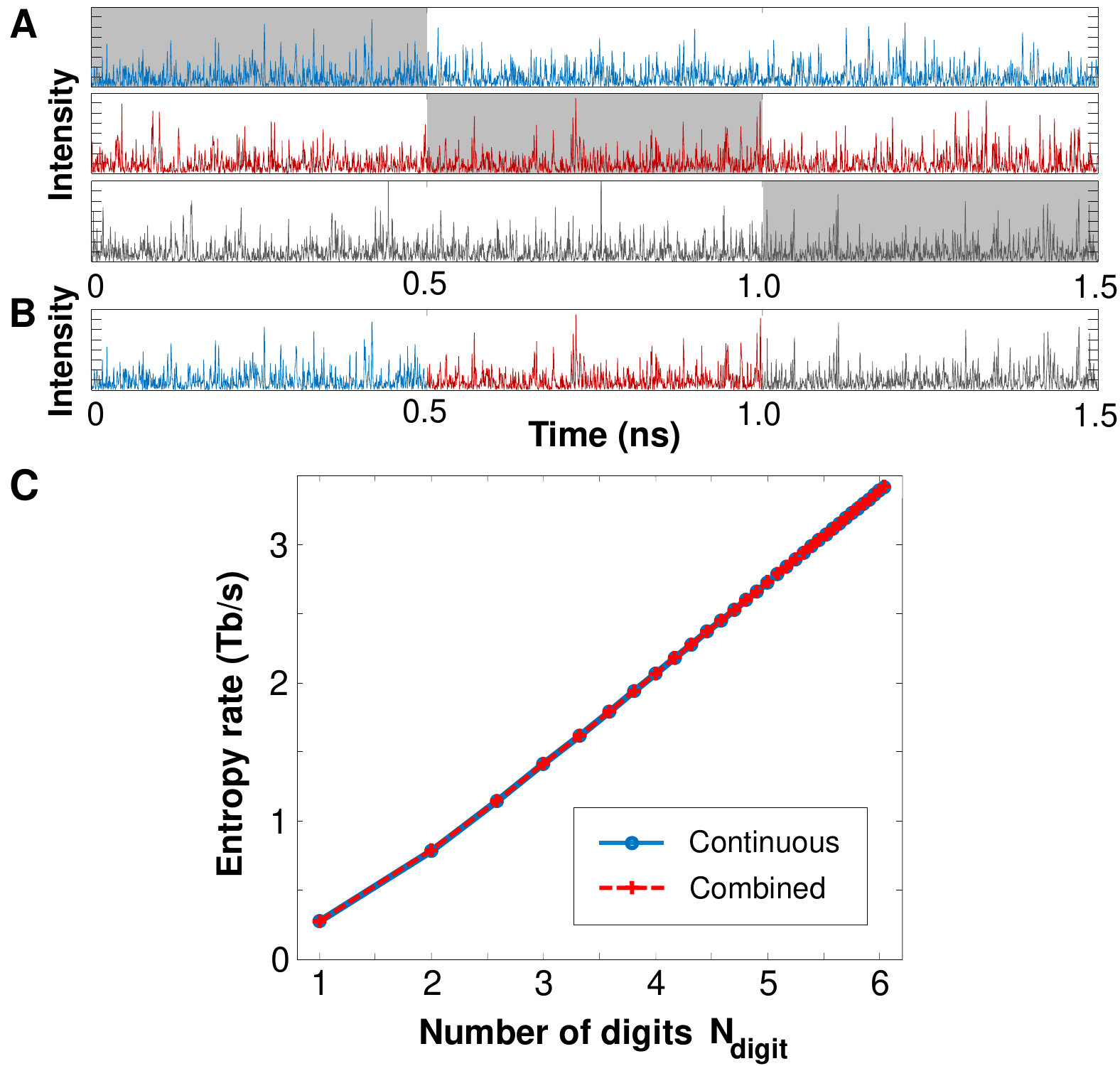}
		\caption{\textbf{Concatenating time traces.}
			(\textbf{A})~Three continuous time traces of the emission intensity in a single spatial channel are generated by simulating the many-mode interference in the laser. The shaded areas represent the temporal range measured by the streak camera.
			(\textbf{B})~A time trace is formed by concatenating the shaded parts of the three traces in (A).
			(\textbf{C})~The Cohen-Procaccia entropy rate estimate of a single 1.5-$\mu$s-long trace as in (A) is equal to that of the concatenated trace as in (B). The sampling period $\tau$ is 1.5 ps, and the embedding dimension $d$ is 2.}
		\label{figS14}
	\end{figure}

Since the temporal measurement range of our streak camera is limited, it is impossible to measure a long continuous time trace. Instead we make separate measurements and concatenate the time traces. Since this process could potentially increase the randomness, we check numerically whether the entropy generation rate is changed by it.

    % FIGURE S15
	\begin{figure*}[t]
		\centering
		\includegraphics[width = 0.8\linewidth]{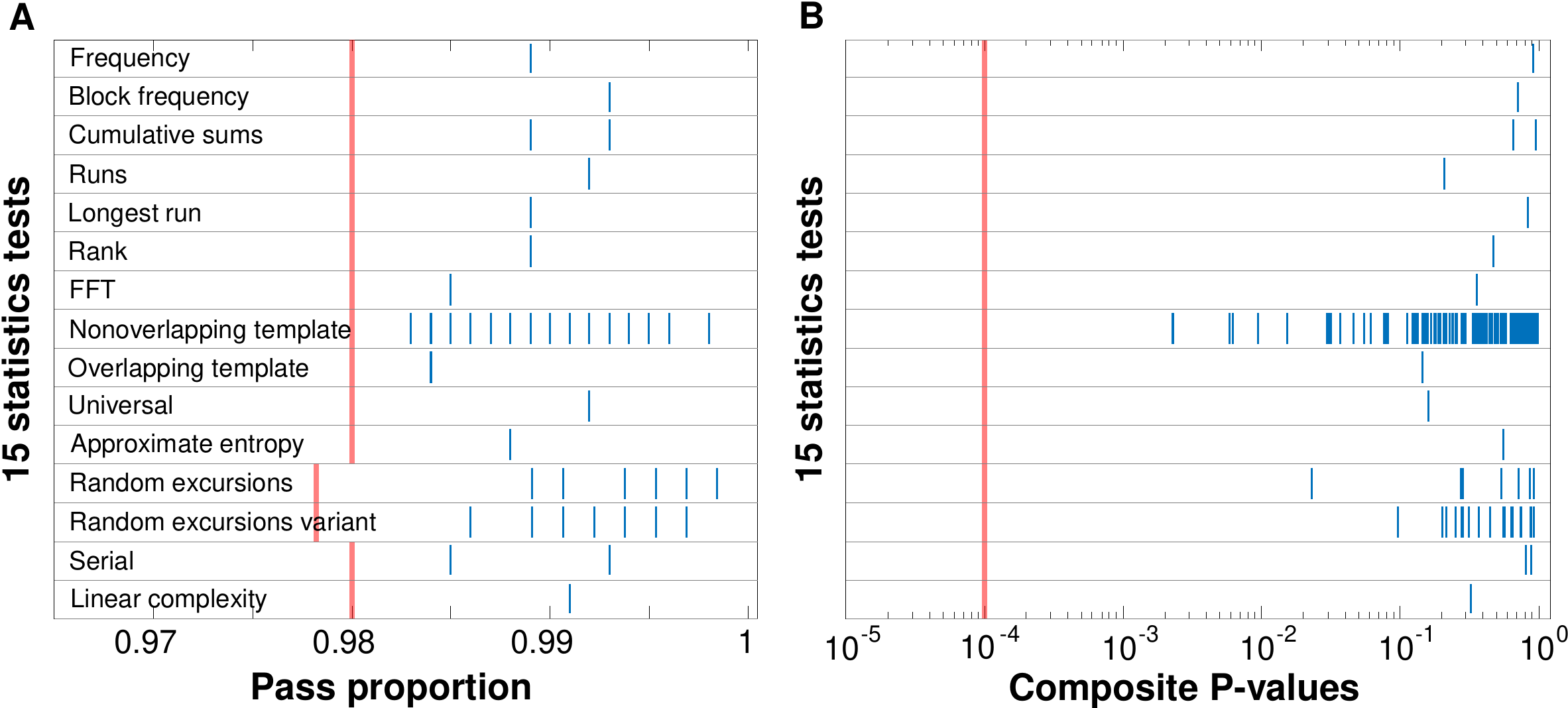}
		\caption{\textbf{NIST SP800-22 statistical test results for a single channel.}
			(\textbf{A})~Pass proportions and
			(\textbf{B})~composite P-values of 15 kinds of statistical tests for random bits generated experimentally in a single channel (66th channel of Fig.~3E\&F). Multiple (short, vertical) blue bars represent the subtests for each kind of statistical test. The red lines denote the pass criteria recommended by NIST.}
		\label{figS15}
	\end{figure*}

Using Eq.~(\ref{eqS11}) to simulate the many-mode interference, we obtain a 1.5-$\mu$s-long time trace of the emission intensity in a single spatial channel. We repeat this process with different random phases for each mode, and obtain 3000 time traces. Short segments of three such traces are shown in Fig.~\ref{figS14}A. Then we extract 500-ps-long segments in consecutive time windows from each trace, and concatenate these segments together for a trace of length 1.5 $\mu$s (Fig.~\ref{figS14}B). We calculate the Cohen-Procaccia entropy rate of the original trace and the concatenated trace. Fig.~\ref{figS14}C shows complete agreement of the two curves, indicating that the process of concatenating windows from different time traces does not affect entropy generation. We also concatenate the same time windows of different traces, and the Cohen-Procaccia entropy rate is the same as well. As the original time trace already reaches the maximal entropy rate (Fig.~4A), concatenating separate traces cannot increase the entropy rate any further.\\

\subsection{Random bit evaluation}

\subsubsection{NIST tests}

We evaluate the quality of the generated random bits with the NIST SP 800-22 test suite. Figure~\ref{figS15} shows the results of NIST tests conducted on a single bit stream containing 1000 sequences with 1 Mbit length per sequence (in total 1 Gbit). The pass proportions and the composite P-values are all above the criteria recommended by NIST, indicating that the quality of the experimentally generated random bit stream is acceptable.

Figure 3E shows that 75$\%$ of the parallel random bit streams can pass the entire NIST test. Considering the statistical nature of the NIST tests, the pass rate was previously evaluated for pseudo-random number generators and physical random number generators. In Ref.~\cite{yamaguchi2010pass}, the NIST test was applied to 100 different sample data sets (each set consisting of 1000 segments of 1 Mbit length), and a pass rate of 41\%$-$56\% was obtained for various representative pseudo-random bit generators. In Ref.~\cite{lihua2014study}, a pass rate of 59\%$-$71\% was reported for some well-known pseudo-random number generation algorithms including those recommended by NIST. In Ref.~\cite{shinohara2017chaotic}, a pass rate between 65\% and 75\% was obtained for a physical RBG based on a chaotic laser. By accounting for the correlations of the sub-tests included in the NIST test suite, the upper bound of the pass rate was estimated to be 80.99\% \cite{lihua2014study}. Compared to the pass rates in these prior studies, our pass rate of 75\% for all channels (Fig.~3E) is considered acceptable for reliable random bit generators.

    % FIGURE S16
	\begin{figure*}[t]
		\centering
		\includegraphics[width = 0.9\linewidth]{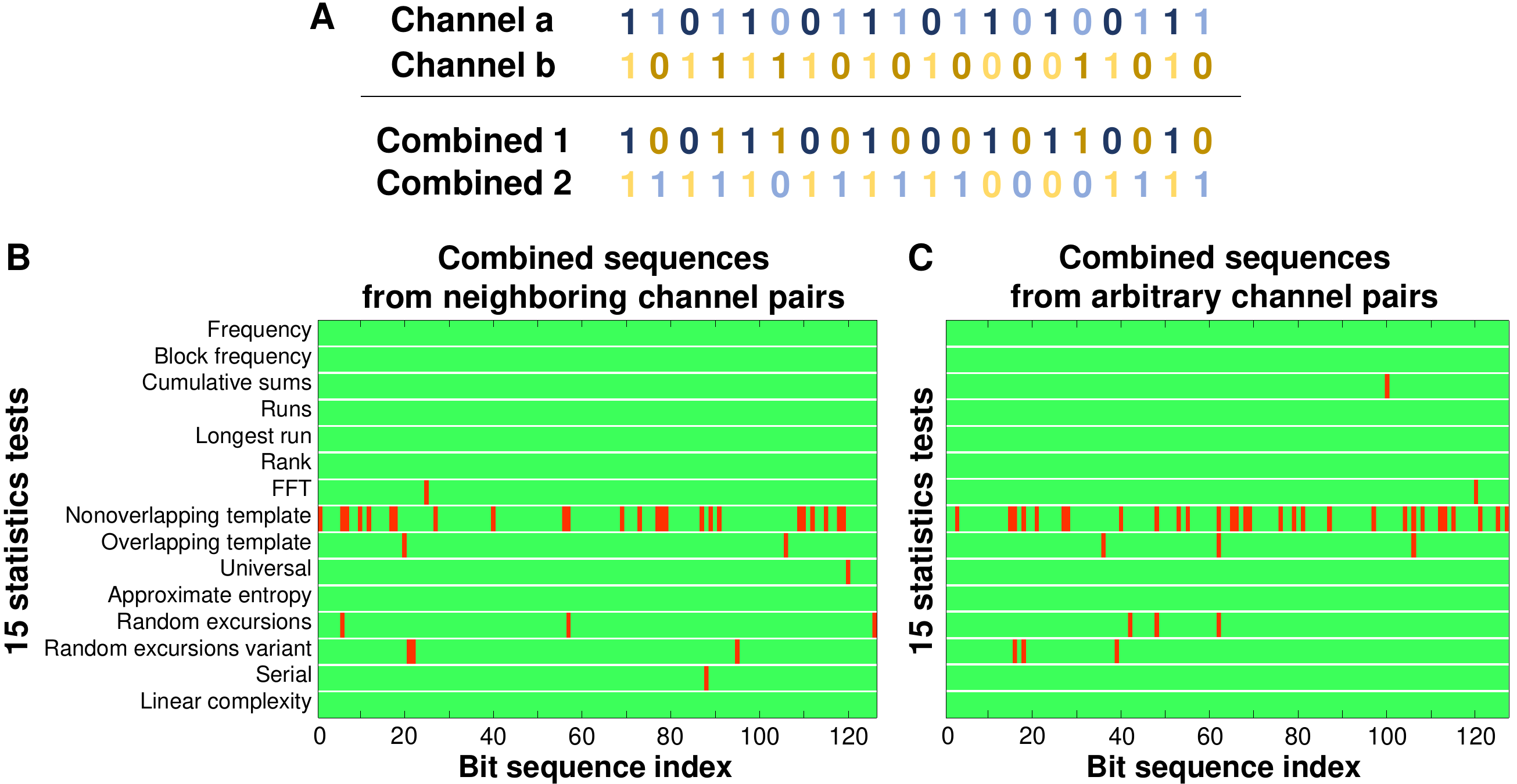}
		\caption{\textbf{NIST SP800-22 statistical test results for combined bit sequences.}
			(\textbf{A})~Schematic of combining two bit streams from channels a \& b. The odd (even)-indexed bits from channel a is combined with the even (odd)-indexed bits from channel b.
			(\textbf{B})~The NIST test results for combined bit sequences created from neighboring spatial locations. The random bit streams from two adjacent channels are combined to produce two new bit sequences. In total 126 combined bit sequences are created from 126 original bit streams. 92 of them pass all NIST tests, yielding a pass rate of 73\%. 
			(\textbf{C})~The test result for combined bit sequences of randoms pairs of channels. 127 pairs are chosen randomly among the 127 channels, and a new bit sequence is constructed from each pair. 92 out of 127 pass all NIST tests, yielding a pass rate of 72\%.}
		\label{figS16}
	\end{figure*}

To verify the independence of the random bit streams from different spatial locations, we combine the bit streams that are generated in parallel, using the procedure illustrated in Fig.~\ref{figS16}A. We first test pairs from spatially adjacent locations which have potentially stronger correlations than non-neighboring pairs. Among all bit streams (after XOR operation on two spatial channels), the even (or odd) bits from the $(2m-1)$-th bit stream and odd (or even) bits from the $2m$-th bit stream are combined to create a new bit sequence with 1 Gbit ($m = 1,\cdots,63$). Out of 126 combined bit sequences, 92 passed all NIST tests, yielding a pass rate of 73\%. Next, to exclude long-range correlations, we combine bit streams that are not necessarily neighbors, by randomly picking two and combining the even bits from one with the odd bits from the other. Out of 127 combined bit sequences, 92 passed all NIST tests, yielding a pass rate of 72\%. All the pass rates are within the acceptable range for reliable RBG. These test results demonstrated the validity and independence of parallel random bit streams generated by our method.

    % FIGURE S17
	\begin{figure*}[t]
		\centering
		\includegraphics[width = 0.9\linewidth]{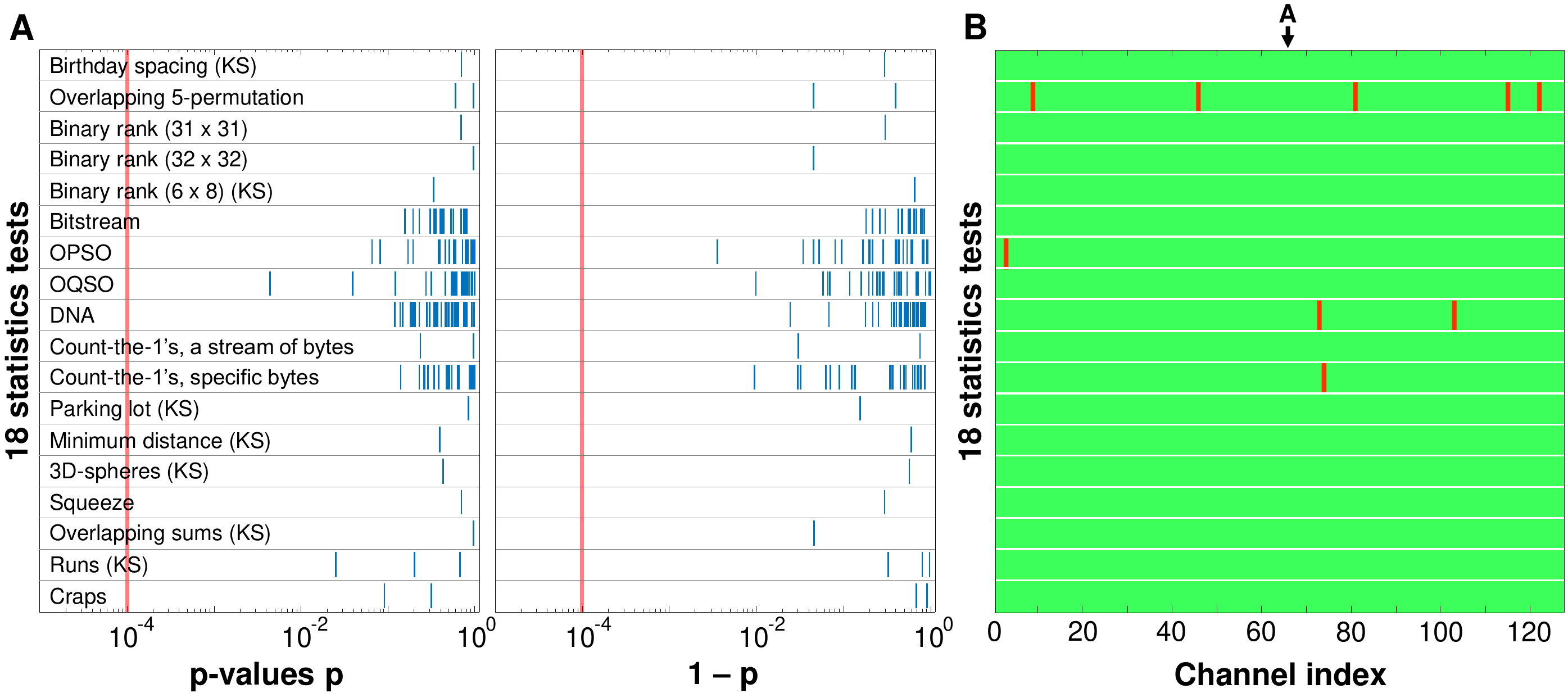}
		\caption{\textbf{Diehard statistical test results.}
			(\textbf{A})~p-values for 18 kinds of statistical tests are obtained with 100 Mbit from a single channel. Each blue (short, vertical) bar represents one p-value. (KS) denotes that the Kolmogorov-Smirnov test is performed to obtain the composite p-value. The red lines denote the minimum and maximum of the acceptable range of p-values: 0.0001 (left panel for $p$) and 0.9999 (right panel for $1-p$).
			(\textbf{B})~The test results for all channels. Red color indicates that the random bits fail the test. The black arrow denotes the channel (66th) used for (A). 118 out of 127 bit streams pass all the tests, yielding a pass rate of 93\%.}
		\label{figS17}
	\end{figure*}
	
\subsubsection{Diehard tests}

We conduct the Diehard tests to further assess the quality of parallel random bit streams. The Diehard test suite consists of 18 kinds of statistical tests~\cite{marsaglia1996diehard}. Each test returns a single or multiple p-values. Some of the tests return a large number of p-values, and a composite p-value is calculated by the Kolmogorov-Smirnov (KS) test to determine if the p-values are uniformly distributed in [0,1). A random bit stream of bad quality returns p-values very close to 0 or 1. The entire test suite passes with a 95\% confidence interval for p-values between 0.0001 and 0.9999~\cite{tsoi2007high}. Figure~\ref{figS17}A shows the Diehard test results for 100 Mbit from a single channel. The p-values from all the statistical tests are within the interval [0.0001, 0.9999], thus the entire test suite is passed. 

	% Table S1
	\begin{table*}[t]
	    \centering
    	\begin{tabular}{ |c||c|c|c|c|c|c|c|c|c|c||c|  }
        \hline
        Test set & 1 & 2 & 3 & 4 & 5 & 6 & 7 & 8 & 9 & 10 & Average \\
        \hline
        Pass rate (\%) & 94 & 94 & 91 & 93 & 92 & 94 & 93 & 91 & 92 & 95 & 93 $\pm$ 1 \\
        \hline
        \end{tabular}
        \caption{\textbf{Pass rate of the Diehard tests.}
        The percentage of parallel bit streams from 127 channels, each 100 Mbit long, that completely pass the Diehard tests. The same tests are performed over 10 independent sets of random bits generated by our laser. In total, 127 $\times$ 10 $\times$ 100 Mbit are tested.}
        \label{tableS1}
    \end{table*}
    
Figure~\ref{figS17}B shows the test results for all channels. Considering the statistical nature of the Diehard test, we evaluate the pass rate over all bit streams. Among the 127 bit streams, 118 bit streams completely pass the Diehard test, yielding a pass rate of 93\%. 

We repeat the tests for 10 different sets of data, and the pass rates are listed in Table~\ref{tableS1}. They range from 91\% to 95\%, and the average pass rate is $93 \pm 1$\%. As a reference, we repeat the tests with random bits generated by one of the widely used pseudo-RBG algorithms, Mersenne-Twister. With the same amount of random bits, the average pass rate over 10 tests is $92 \pm 2$\%. It is very close to the pass rate of all bit streams generated by our laser, thus certifying the randomness of our parallel RBG.

% end of document
\end{document}